%% file: main.tex
\pdfoutput=1

\documentclass[11pt]{article}

\usepackage{xcolor}

\usepackage{acl}

\usepackage{times}
\usepackage{latexsym}
\usepackage{graphicx}
\usepackage{amsmath}
\usepackage{booktabs}
\usepackage{colortbl}
\usepackage{multirow}
\usepackage{pifont}
\usepackage{enumitem}
\usepackage{listings}
\usepackage[most]{tcolorbox}
\usepackage{subcaption}
\usepackage{afterpage}
\usepackage{xfp}

\newcommand{\f}[1]{\cellcolor{teal!\fpeval{round(70*#1)}}#1}
\newcommand{\ov}[1]{\cellcolor{teal!\fpeval{round(70*#1/13.21)}}#1}
\newcommand{\err}[1]{\cellcolor{orange!\fpeval{round(70*#1/0.12)}}#1}

\definecolor{userpromptbg}{HTML}{eff6fe}
\definecolor{userpromptborder}{HTML}{6c8ebf}
\definecolor{userprompttitle}{HTML}{aec7e8}
\definecolor{userprompttitletext}{HTML}{2c3e50}

\newtcolorbox{examplebox}[1][]{
    listing only,
    enhanced,
    colback=userpromptbg,
    colframe=userpromptborder,
    colbacktitle=userprompttitle,
    coltitle=userprompttitletext,
    arc=2mm,
    boxrule=1pt,
    left=2mm,
    right=2mm,
    top=2mm,
    bottom=2mm,
    fonttitle=\bfseries\small,
    title=#1,
    toptitle=0.5mm,
    bottomtitle=0.5mm,
    listing options={
        basicstyle=\ttfamily\tiny,
        breaklines=true,
        columns=flexible,
        keepspaces=true,
    },
    drop shadow,
}

\usepackage[T1]{fontenc}

\usepackage[utf8]{inputenc}

\usepackage{microtype}

\title{Loci Similes: A Benchmark for Extracting Intertextualities  \\ in Latin Literature}

\author{Julian Schelb$^\dagger$, Michael Wittweiler$^\diamond$, \textbf{Marie Revellio$^{\ddagger*}$},\\
    \textbf{Barbara Feichtinger$^\ddagger$, \and Andreas Spitz$^{\dagger*}$} \\
  $^\dagger$Department of Computer and Information Science, University of Konstanz \\
  $^\ddagger$Department of Latin Philology, University of Konstanz \\
  $^\diamond$Institute of Archaeology, Classical Philology and Ancient Studies, University of Zurich \\
  \texttt{\{firstname.lastname\}@uni-konstanz.de} \\
  \texttt{\{firstname.lastname\}@uzh.ch}}

\begin{document}
\maketitle

\def\thefootnote{*}\footnotetext{These authors contributed equally as project lead}\def\thefootnote{\arabic{footnote}}

\begin{abstract}
Tracing connections between historical texts is an important part of intertextual research, enabling scholars to reconstruct the \textit{virtual library} of a writer and identify the sources influencing their creative process.  These intertextual references manifest in diverse forms, ranging from direct verbatim quotations to subtle allusions and paraphrases disguised by morphological variation.
Language models offer a promising path forward due to their ability to capture semantic similarity beyond lexical overlap. However, the development of new methods for this task is held back by the scarcity of standardized benchmarks and easy-to-use datasets.
We address this gap by introducing \textit{Loci Similes}, a benchmark for Latin intertextuality detection comprising a curated dataset of $\sim$176k text segments and 1,490 expert-verified intertextual references, including 945 from an existing dataset. Using this data, we establish baselines for retrieval and classification of intertextualities with both lexical methods and pretrained encoder language models.
\end{abstract}

\input{sections/01_introduction.tex}

\input{sections/02_related_work}

\input{sections/03_dataset}

\input{sections/04_framework}

\input{sections/07_results}

\input{sections/07b_error_analysis}

\input{sections/08_discussion}

\bibliography{
    bibliography/anthology,
    bibliography/references,
    bibliography/models,
    bibliography/related_work}
\bibliographystyle{acl_natbib}

\appendix
\input{sections/appendix/09_sources}
\input{sections/appendix/10_setup}
\input{sections/appendix/11_results}
\input{sections/appendix/11b_error_analysis}
\input{sections/appendix/12_framework}
\input{sections/appendix/13_scalability}

\end{document}

%% file: sections/01_introduction.tex
\section{Introduction}
\label{sec:intro}
Identifying intertextual references between documents is an important task in classical philology, as it reveals how later works engage with earlier texts and traditions.
For centuries, scholars detected intertextual references by relying on memory and the manual collation of \textit{Loci Similes}, i.e., parallel passages that exhibit lexical, semantic, or thematic resemblance. Although digitization has augmented this process through lexical search tools, most approaches still depend on exact n-gram matching or heuristic filtering \citep{DBLP:journals/dhq/SchroppKRF24}. This limits discovery rates in ancient texts, where intertextuality not only manifests itself as verbal quotation, but also as subtle allusion, paraphrase, or thematic variation \citep{DBLP:conf/latech/ManjavacasLK19, gong_augmented_2025}, often complicated by orthographic volatility \citep{Miller2025Alignment}.

\begin{figure}[t]
    \vspace{0.1cm}
    \centering
    \begin{examplebox}[Example of Historical Text Reuse]
        \small
        \textbf{Source: Virgil, \textit{Aeneid} 2.774} \\
        \textit{Context:} Aeneas is terrified when the ghost of his wife Creusa appears to him during the burning of Troy.
        \begin{quote}
            ``... obstipui, steteruntque comae et \textbf{uox faucibus haesit}.'' \\
            \textcolor{gray}{\textit{(... I was stupefied, my hair stood on end, and my \textbf{voice stuck in my throat}.)}}
        \end{quote}
        
        \vspace{0.0cm}
        
        \textbf{Reuse: Jerome, \textit{Epistula} 130.5.5} \\
        \textit{Context:} A family's shocked reaction to Demetrias's vow of Christian virginity.
        \begin{quote}
            ``\textbf{Haesit uox faucibus} et inter ruborem atque pallorem metumque ac laetitiam cogitationes uariae mutabantur.'' \\
            \textcolor{gray}{\textit{(\textbf{The voice stuck in their throat}, and between blushing and pallor, fear and joy manifold thoughts kept shifting.)}}
        \end{quote}
    \end{examplebox}
    \vspace{-0.2cm}
    \caption{\textbf{Example of intertextual reference.} Reuse of a classic Virgilian phrase for speechlessness by Jerome. While retaining the semantic core, the author alters the word order to adapt the expression to a different context.}
    \label{fig:intertextual_example}
    \vspace{-0.5cm}
\end{figure}

Recovering such text reuse is not merely a matter of identifying sources. It facilitates research on broader cultural-historical phenomena \citep{tangherlini2024travels}. It supports work on reception and cultural hybridization in Late Antiquity, where pagan texts persist as the rhetorical substrate of elite writing while being recontextualized within emerging Christian discourse. Classical forms remain recognizable even as their functions shift toward Christian meaning-making, visible in computational stylometry \citep{gorman2016approaching, DBLP:journals/corr/abs-2109-00601} and semantic motifs. Citation patterns suggest how Christian authors do not abandon classical texts and their cultural contexts, but reuse their language, redirecting its meanings within Christian frameworks.

A case in point is the Church Father Jerome. Recent digital-hermeneutic studies have begun to map his ``micro-quotations'' \citep{DBLP:journals/dhq/SchroppKRF24}, yet the semantic breadth of his reuse remains a challenge. When Jerome alludes to the Augustan poet Virgil, he often retains the semantic core of a hexameter verse while altering its word order or syntax to suit his Christian prose context. In Jerome's writings, the general tension between his pagan \textit{paideia} and Christian discourse emerges with particular clarity in the details of how he quotes classical pagan sources and adapts them for his own texts (see Figure \ref{fig:intertextual_example}). Since canonical authors such as Virgil were deeply embedded in the educational curriculum, quoting them often served as a shorthand for shared cultural memory. Late Antique Christian writers like Jerome inherit this repertoire, but their reuse frequently reframes pagan language within Christian contexts, making citation patterns a measurable trace of shifting cultural authority.

Beyond extraction, systematically mapping these references allows scholars to reconstruct the ``virtual library'' available to an author and to gain insight into which sources most strongly shaped their writing. Analyzing how these references cluster, from explicit citation to subtle echo, further helps refine theoretical definitions of intertextuality and probe the rhetorical motivations behind text reuse. Although some resources exist~\citep{DBLP:conf/naacl/BurnsBLCD21,toyin-etal-2026-gretino}, language-model-based detection of intertextual references remains constrained by a lack of standardized benchmarks and datasets.

In this paper, we take a step towards addressing this gap by introducing \textit{Loci Similes}, a benchmark designed to enable researchers to systematically compare and evaluate computational approaches for intertextuality detection.\footnote{Dataset, fine-tuned models, and code are available at \href{https://github.com/julianschelb/locisimiles}{https://github.com/julianschelb/locisimiles}.}
We contribute:

\begin{itemize}[leftmargin=10pt, itemsep=3pt, parsep=0pt, topsep=3pt]
    \item \textbf{Curated benchmark dataset} of $\sim$176k Latin text segments, partitioned into query and source corpora, accompanied by a ground-truth dataset of 1,490 expert-verified intertextual references.
    \item \textbf{Evaluation framework} that aligns more closely with the practical constraints of philological workflows.
    \item \textbf{Baseline results} for retrieval, classification, and a retrieve-and-rerank pipeline, serving as a foundation for future comparisons.
\end{itemize}

%% file: sections/02_related_work.tex
\section{Related Work}
\label{sec:relwork}

\subsection{Intertextuality Detection in Latin}
\label{subsec:related_work_latin}

Work on Latin intertextuality ranges from surface-level matching to increasingly semantic representations.
\citet{revellio2022zitate} developed a rule-based approach operating directly on surface strings to uncover Virgilian references in Jerome's letters, and \citet{DBLP:journals/dhq/SchroppKRF24} refined this with n-gram matching pipelines and cascaded filters to detect very short references. Such methods capture verbatim and near-verbatim reuse well, but miss semantically related passages with little shared vocabulary.

Word-level embeddings instead capture semantic similarity beyond lexical overlap. \citet{DBLP:conf/naacl/BurnsBLCD21} used static Word2Vec models trained on lemmatized text to rank intertextual phrases, evaluated against 945 parallels from Valerius Flaccus' \textit{Argonautica}~\citep{dexter_ldquodatabase_2024}. \citet{DBLP:conf/latech/ManjavacasLK19} approached allusion detection in Latin sermons as information retrieval using fastText embeddings, and \citet{DBLP:conf/chr/ManjavacasKK20} extended this work on the \textit{Patrologia Latina}, modeling lexical and thematic similarity as separate axes.

\citet{gong_augmented_2025} replaced static vectors with contextual LatinBERT~\citep{DBLP:journals/corr/abs-2009-10053} embeddings, retrieving a target word's nearest neighbors by cosine similarity to identify allusions in Lucan's \textit{Pharsalia}. \citet{DBLP:conf/ranlp/Riemenschneider23} applied sentence-transformer models~\citep{DBLP:conf/emnlp/ReimersG19} to ancient languages, introducing SPhilBERTa, a trilingual model built on PhilBERTa~\citep{DBLP:conf/acl/Riemenschneider23} and fine-tuned on parallel sentences in Ancient Greek, Latin, and English. More recently, \citet{toyin-etal-2026-gretino} introduced Gretino, a benchmark for semantic retrieval in Latin, Ancient Greek, and cross-lingual settings. Unlike these resources, \textit{Loci Similes} labels positive references by reference type, allowing verbatim references and allusions to be evaluated separately.

\subsection{Text Reuse in Other Languages}
\label{subsec:related_work_other}

Beyond Latin, \citet{DBLP:journals/coling/KuznetsovBEG22} model intertextual relations as a graph, \citet{DBLP:journals/corr/abs-2410-15145} use large language models to extract asymmetric intertextuality, and \citet{DBLP:conf/starsem/MacLaughlinXS21} show that contextual models outperform lexical baselines. \citet{DBLP:journals/corr/abs-2302-04084} built a tool for exploring reuse in Early Modern British texts, whose pipeline~\citep{DBLP:journals/ijdsa/MahadevanMMT25} indexes billions of reuse pairs.

Specialized approaches address low-resource languages. \citet{Miller2025Alignment} enhanced reuse detection in Hebrew and Aramaic manuscripts using fastText embeddings, and \citet{gorman2016approaching} modeled the authorial style in Ancient Greek through unsupervised clustering. \citet{DBLP:journals/talip/SharjeelMNNR23} addressed cross-lingual reuse in English--Urdu using machine translation.

\subsection{Further Related Tasks}
\label{subsec:related_work_similar_tasks}

Intertextuality detection is related to quote detection~\citep{DBLP:conf/nodalida/JanickiKM23} and paraphrase identification~\citep{DBLP:journals/es/VrbanecM23}. However, quote detection targets speech explicitly attributed to a speaker~\citep{DBLP:conf/konvens/PetersenFreyB24, DBLP:conf/lrec/ZhangL22}, and paraphrase detection focuses on semantic equivalence~\citep{DBLP:conf/emnlp/WahleRKG22}. Our task differs in requiring the capture of unmarked allusions and recontextualized echoes at the segment level across unrelated documents.

%% file: sections/03_dataset.tex
\begin{figure}[t]
    \centering
    \includegraphics[width=1.0\columnwidth]{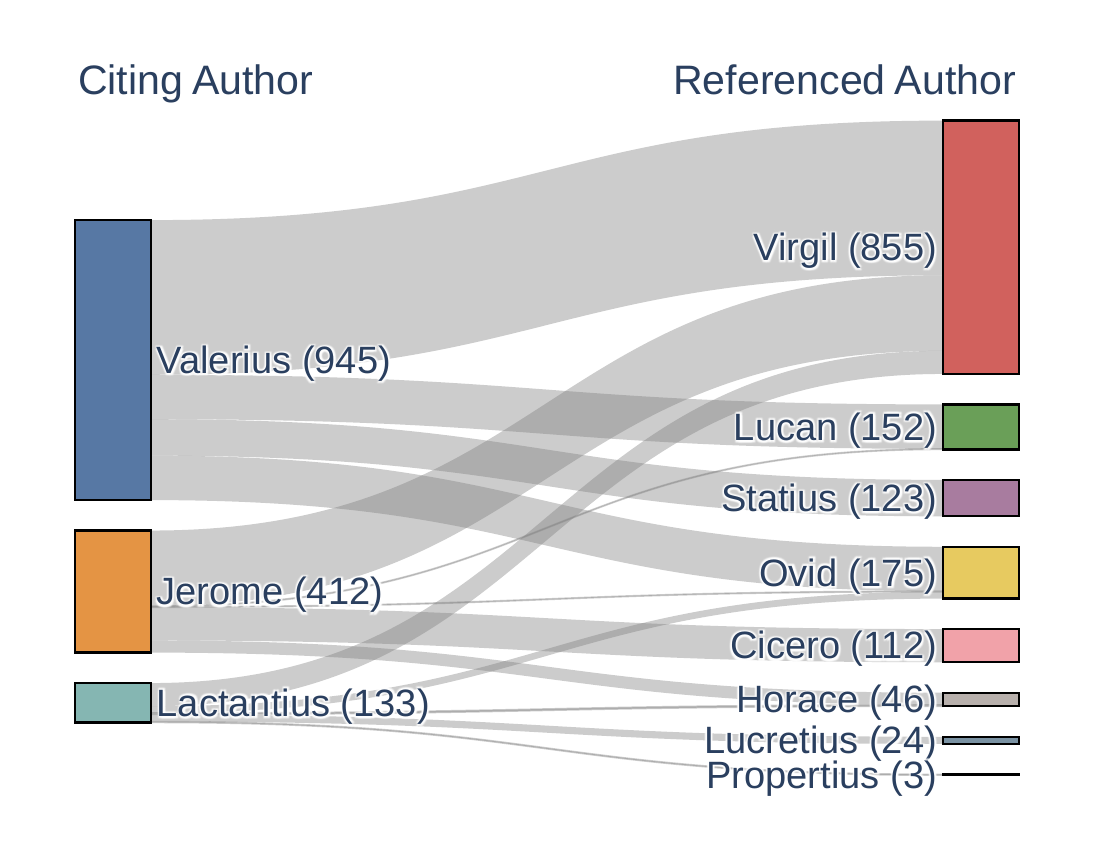}
    \caption{\textbf{Distribution of confirmed intertextual references.} Manually verified intertextual references in the annotated dataset by citing author (Jerome, Lactantius, and Valerius Flaccus) and source author (including Virgil, Cicero, and others).}
    \label{fig:textreuse}
    \vspace{-0.5cm}
\end{figure}
\begin{table}[t]
\centering
\resizebox{\columnwidth}{!}{%
\begin{tabular}{lrrrrr}
\toprule
\textbf{Author} & \textbf{Segments} & \textbf{Avg. Tokens} & \textbf{Min} & \textbf{Max} & \textbf{Std. Dev.} \\
\midrule
\multicolumn{6}{c}{\textbf{Query Corpus}} \\
\midrule
Jerome & 74,672 & 31.13 & 1 & 617 & 22.32 \\
Lactantius & 8,444 & 28.53 & 1 & 369 & 19.48 \\
Valerius & 2,138 & 29.60 & 1 & 145 & 18.10 \\
\midrule
\textit{\textbf{Total / Avg.}} & \textit{\textbf{85,254}} & \textit{\textbf{30.83}} & \textit{\textbf{1}} & \textit{\textbf{617}} & \textit{\textbf{21.97}} \\
\midrule
\multicolumn{6}{c}{\textbf{Source Corpus}} \\
\midrule
Cicero & 54,331 & 28.97 & 1 & 1,377 & 25.40 \\
Ovid & 14,096 & 26.51 & 1 & 432 & 19.53 \\
Virgil & 4,861 & 29.65 & 2 & 353 & 19.55 \\
Martial & 4,114 & 26.13 & 2 & 274 & 21.07 \\
Lucan & 2,952 & 31.26 & 2 & 245 & 21.83 \\
Statius & 2,527 & 44.08 & 2 & 412 & 40.22 \\
Horace & 2,353 & 33.60 & 2 & 195 & 24.83 \\
Propertius & 1,889 & 22.18 & 2 & 267 & 17.56 \\
Lucretius & 1,826 & 40.95 & 3 & 256 & 27.40 \\
Catullus & 809 & 27.43 & 2 & 149 & 22.08 \\
Tibullus & 788 & 26.12 & 2 & 251 & 20.70 \\
\midrule
\textit{\textbf{Total / Avg.}} & \textit{\textbf{90,546}} & \textit{\textbf{29.17}} & \textit{\textbf{1}} & \textit{\textbf{1,377}} & \textit{\textbf{24.61}} \\
\bottomrule
\end{tabular}%
}
\caption{\textbf{Corpus statistics by author.} Breakdown of the dataset into \textit{Query} (citing authors) and \textit{Source} (referenced authors) corpora, showing the number of segments and token statistics for each author.}
\vspace{-0.5cm}
\label{tab:corpus_stats}
\end{table}
\section{Dataset Construction}
\label{sec:dataset}

While traditional scholarship has documented numerous Latin intertextual references, computational research remains constrained by the lack of standardized benchmarks. Aggregating these references is non-trivial, as the data is dispersed across commentaries, \textit{indices locorum} (``back-of-the-book indexes''), and philological case studies. To address this gap, we curated a corpus of $\sim$176k text segments from multiple Latin authors (see Table~\ref{tab:corpus_stats}) and a merged ground-truth dataset of 1,490 confirmed intertextual references (see Figure~\ref{fig:textreuse}).

\subsection{Latin Corpus Curation}
\label{subsec:sources}

We compose the corpus of a separate \textit{Query Corpus} and a \textit{Source Corpus}. The former comprises works by the Late Antique authors Jerome and Lactantius, together with Book~1 of Valerius Flaccus' \textit{Argonautica} added through the subset from \citet{DBLP:conf/naacl/BurnsBLCD21}, totaling $\sim$85k text segments. The latter, which serves as the retrieval target, consists of $\sim$91k segments drawn from eleven canonical classical Latin authors, in chronological order: Cicero, Lucretius, Catullus, Virgil, Horace, Tibullus, Propertius, Ovid, Lucan, Martial, and Statius. Text segments correspond to sentences delimited by hard punctuation (periods, exclamation marks, and question marks). A detailed breakdown of segment counts and token statistics for each author is presented in Table~\ref{tab:corpus_stats}. The texts were aggregated from three primary digital repositories: \textit{Corpus Corporum}, the \textit{Tesserae Project}, and the \textit{OpenGreekandLatin Project}. For a complete listing of the specific works, critical editions used, and their respective provenance, see Appendix~\ref{app:sources}. Since we exclusively rely on publicly available texts to compile the corpus, we release the corpus files alongside the ground-truth dataset.

\subsection{Ground Truth Construction}
\label{subsec:ground_truth}

Our ground-truth dataset combines 545 references curated for \textit{Loci Similes} with 945 labeled references from Valerius Flaccus' \textit{Argonautica}~\citep{DBLP:conf/naacl/BurnsBLCD21}, 1,490 in total.

First, we sourced 270 references to Virgil and Cicero in Jerome's \textit{Epistulae} from the dataset\footnote{\href{https://doi.org/10.11588/data/FVCULR}{https://doi.org/10.11588/data/FVCULR}} established by \citet{DBLP:journals/dco/SchroppWKRF24}, adapting these entries to sentence-level granularity and excluding references disputed by our expert annotators. We added 275 further references, identified with the rule-based \textit{n-gram} matching approach of \citet{DBLP:journals/dhq/SchroppKRF24}.\footnote{\href{https://github.com/MWittweiler/citation_analysis}{https://github.com/MWittweiler/citation\_analysis}}
Their pipeline scans for shared, non-contiguous tokens within a fixed window, then refines the candidate pairs through a cascade of filters that remove stopwords, enforce part-of-speech constraints, and exclude generic collocations by embedding similarity. Domain experts then manually evaluated the candidates, retaining only those confirmed as genuine intertextual references.

Second, we incorporated the dataset of \citet{DBLP:conf/naacl/BurnsBLCD21}, comprising 945 known references between Book~1 of the \textit{Argonautica} and four major Latin epics: Virgil's \textit{Aeneid}, Ovid's \textit{Metamorphoses}, Lucan's \textit{Pharsalia}, and Statius' \textit{Thebaid}. These references derive from traditional scholarship, namely the commentaries of \citet{spaltenstein2002commentaire}, \citet{kleywegt2005valerius}, and \citet{zissos2008valerius}.
We aligned their granularity and format with our existing dataset.

Item-level scholarly support extends to nearly the entire benchmark: the 1,215 references adopted from the two existing datasets derive from published commentaries, and 274 of the 275 references added through the n-gram pipeline are likewise attested in published editions or commentaries, so that 1,489 of the 1,490 positive references are documented in prior scholarship. The remaining reference is, to our knowledge, a new discovery and will be described in a separate philological publication. For each reference, we record the publication from which it was adopted, so its scholarly support can be verified directly.

\begin{figure}[t]
    \centering
    \begin{examplebox}[Taxonomy of Intertextual References]
        \small
        \textbf{1. Verbatim Reference (\textit{cit.}) -- Literal Reuse}\\[2pt]
        \textit{Near-verbatim reuse sharing at least two congruent lemmas in a coherent context.}
        \vspace{2pt}
        \begin{itemize}[leftmargin=20pt, noitemsep, topsep=1pt]
            \item \textit{Marked:} Explicitly attributed to a specific author or source (e.g., \textit{ut ait Cicero...} \textrightarrow{} ``as Cicero says'').
            \item \textit{Unmarked:} Silent incorporation of exact text strings from other sources.
            \item \textit{Paraphrased:} Morphological, word-order, or lemma-equivalent variation (\textit{arma uirumque} \textrightarrow{} \textit{armis uirisque}; \textit{cognoscat} \textrightarrow{} \textit{cognoscere}).
        \end{itemize}

        \vspace{0.2cm}

        \textbf{2. Allusion (\textit{cf.}) -- Semantic Similarity}\\[2pt]
        \textit{Thematic or stylistic echo without close lexical overlap; synonyms, paraphrase, or shared imagery instead of shared lemmas.}
        \vspace{2pt}
        \begin{itemize}[leftmargin=20pt, noitemsep, topsep=1pt]
            \item \textit{Single reference:} Shared distinct vocabulary or imagery recontextualized in a new setting (\textit{infans} \textrightarrow{} \textit{puer}; \textit{paruulus} \textrightarrow{} \textit{paruus}).
            \item \textit{Systemic:} Broad stylistic, rhythmic, or thematic imitation on multiple occasions in the document.
        \end{itemize}
    \end{examplebox}
    \caption{\textbf{Annotation scheme.} Following \citet{DBLP:journals/dco/SchroppWKRF24}, we adopt a two-way reference-type scheme based on \textit{Verbatim References} (\textit{cit.}) and \textit{Allusions} (\textit{cf.}).}
    \label{fig:intertext_taxonomy}
    \vspace{-0.5cm}
\end{figure}

\subsection{Annotation Process}
\label{subsec:annotation}

Constructing the ground truth involved two decisions: establishing whether a candidate pair constitutes an intertextual reference and, if so, assigning its reference type.

\paragraph{Establishing a reference.} The pipeline's filters removed many false positives, but numerous candidates still involved common collocations (e.g., \textit{puncto temporis} ``in a moment'') and required manual exclusion. Four experts in Latin literature (two pre-PhD, two post-PhD) therefore assessed each of the 275 n-gram-derived candidates against the relevant scholarship, confirming references by consensus rather than by independent parallel annotation. Their assessment was guided by three criteria:

\begin{enumerate}[leftmargin=15pt, itemsep=0.5pt,topsep=5pt]
    \item \textbf{Use of Uncommon Vocabulary:} If the lexical overlap consists of rare or marked words, the likelihood increases that these elements were deliberately borrowed.
    \item \textbf{Attested Frequency:} The \textit{Library of Latin Texts} corpus database\footnote{\href{http://clt.brepolis.net/llta/Search}{http://clt.brepolis.net/llta/Search}} served to determine the frequency of the overlapping expression in Latin literature. Expressions appearing exclusively in the candidate source and target passages were treated as strong indicators of a unique intertextual relationship.
    \item \textbf{Conduit Function:} Most importantly, if the cited expression contributes semantic, rhetorical, or cultural information from the source text that cannot be derived from the target passage in isolation, thereby implicitly enriching the interpretation, it is viewed as intertextual.
\end{enumerate}

\paragraph{Assigning a reference type.} The reference type is assigned with the operational criterion of \citet{DBLP:journals/dco/SchroppWKRF24}. An established reference is labeled a \textit{verbatim reference} if it contains at least two matching lemmas in close proximity within text segments delimited by hard punctuation. Close proximity permits at most two intervening words after excluding stopwords such as \textit{et}, \textit{sic}, and \textit{non}. Established references that do not meet these criteria are labeled \textit{allusions}. Given the lemma matches, distances, and punctuation, the classification is deterministic and requires no assessment of thematic, stylistic, or poetic similarity. We applied the same scheme across all sources: reusing the reference-type labels for the 270 references from \citet{DBLP:journals/dco/SchroppWKRF24} and manually annotating the 275 newly added references and the 945 from \citet{DBLP:conf/naacl/BurnsBLCD21}.
Figure~\ref{fig:intertext_taxonomy} describes the two positive reference types. Together with unrelated pairs, this yields a multiclass labeling into \textit{no match}, \textit{verbatim reference}, and \textit{allusion}, enabling finer evaluation, since explicit lexical reuse, looser allusive correspondences, and pairs labeled \textit{no match} pose substantially different retrieval and classification problems.

We release the complete dataset under the Apache~2.0 license as a Hugging Face collection comprising the source corpus, the query texts, and the annotated intertextual references.\footnote{\href{https://hf.co/collections/julian-schelb/datasets-for-latin-intertextuality-search}{https://hf.co/collections/julian-schelb/datasets-for-latin-intertextuality-search}} The fine-tuned retrieval and classification models reported in Section~\ref{sec:experiments} are released as a companion Hugging Face collection.\footnote{\href{https://hf.co/collections/julian-schelb/models-for-latin-intertextuality-search}{https://hf.co/collections/julian-schelb/models-for-latin-intertextuality-search}}

%% file: sections/04_framework.tex
\begin{figure}[!t]
\centering
\includegraphics[width=\linewidth]{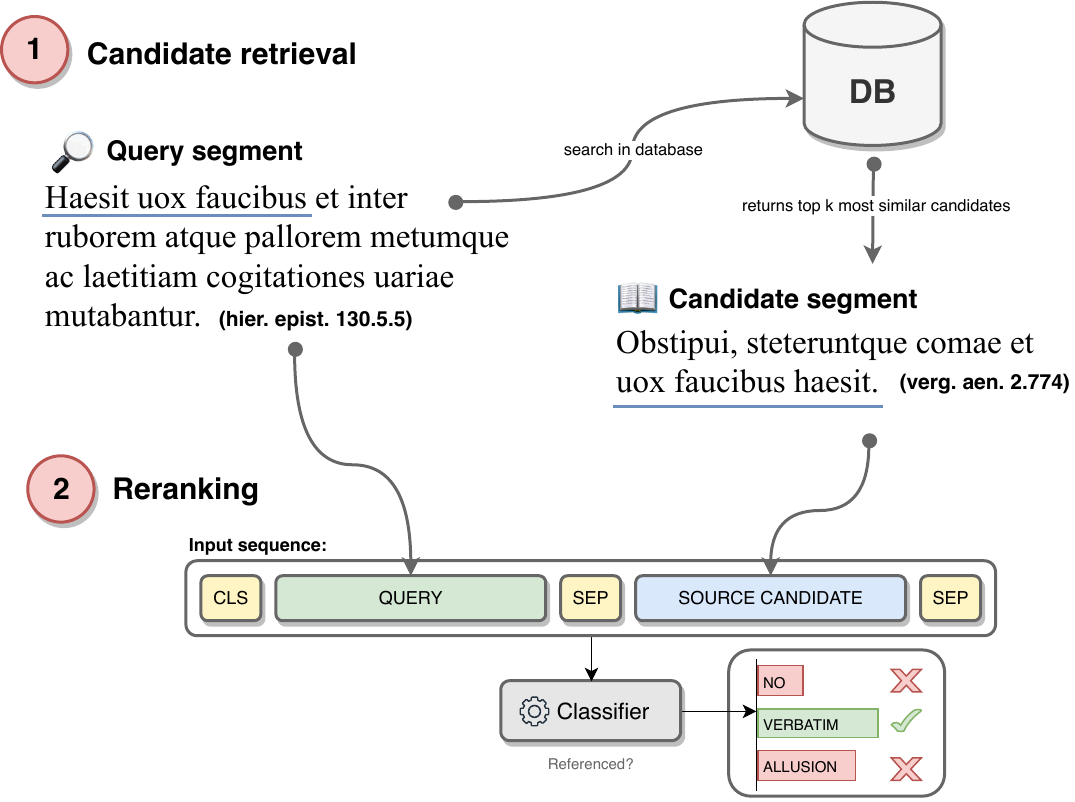}
\caption{\textbf{Two-stage experimental setup.} \textit{Stage 1:} The input text segment acts as a query to retrieve potential candidates from the database. \textit{Stage 2:} To verify the reference, the query and source candidate are concatenated into a single input sequence and passed to a three-class cross-encoder (\textit{no match}, \textit{cit.}, \textit{cf.}).}
\label{fig:two_stage_pipeline}
\vspace{-0.3cm}
\end{figure}

\begin{figure*}[t]
\centering
\small
\begin{minipage}[t]{0.49\textwidth}
\centering
\resizebox{\linewidth}{!}{%
\begin{tabular}{lrrrrrr}
\toprule
\multirow{2}{*}{\textbf{Model}} & \multicolumn{3}{c}{\textbf{Verbatim (cit.) Rec.}} & \multicolumn{3}{c}{\textbf{Allusion (cf.) Rec.}} \\
\cmidrule(lr){2-4}\cmidrule(lr){5-7}
 & \textbf{10} & \textbf{100} & \textbf{1000} & \textbf{10} & \textbf{100} & \textbf{1000} \\
\midrule
\textbf{TF-IDF (surf. 1g)} & 60.63 & 75.39 & 87.55 & 8.52 & 23.59 & 50.97 \\
\textbf{TF-IDF (lem. 1g)} & 55.47 & 72.28 & 90.09 & 8.14 & 20.79 & 56.61 \\
\textbf{TF-IDF (lem. 1+2g)} & 58.19 & 74.03 & \textbf{90.78} & 7.90 & 25.70 & \textbf{60.74} \\
\textbf{BM25 (surf. 1g)} & \textbf{67.70} & 75.64 & 87.01 & \textbf{12.22} & 28.40 & 50.29 \\
\textbf{BM25 (lem. 1g)} & 63.97 & \textbf{77.62} & 90.08 & 11.96 & \textbf{29.43} & 57.92 \\
\midrule
\textbf{BGE-M3} & 50.06 & 63.50 & 78.20 & 7.52 & 20.62 & 44.01 \\
\textbf{E5-large} & 49.93 & 63.06 & 79.54 & 7.34 & 18.25 & 48.42 \\
\textbf{E5-base} & 48.73 & 61.32 & 77.45 & 5.61 & 16.10 & 43.82 \\
\textbf{E5-small} & 44.17 & 58.03 & 74.63 & 5.68 & 13.99 & 34.56 \\
\textbf{Granite-278m} & 43.09 & 54.18 & 69.43 & 3.36 & 12.31 & 36.10 \\
\textbf{Granite-107m} & 31.35 & 41.04 & 57.56 & 2.27 & 6.68 & 27.28 \\
\textbf{SPhilBERTa} & 40.94 & 56.33 & 73.87 & 6.19 & 19.27 & 48.29 \\
\midrule
\textbf{LatinBERT (WE)} & 56.70 & 68.84 & 82.83 & 10.03 & 28.14 & 52.70 \\
\textbf{Word2Vec (WE, 2g)} & 14.65 & 30.62 & 60.86 & 0.51 & 6.83 & 25.33 \\
\bottomrule
\end{tabular}
}
\vspace{0.35em}
\captionof{table}{\textbf{Performance comparison of retrieval approaches.}
All values are $\times100$, macro-averaged over query segments and averaged over the 5 folds. Lexical approaches (TF-IDF, BM25) use surface tokens (\textit{surf}) or lemmas
(\textit{lem}) with unigrams (\textit{1g}) or unigrams+bigrams (\textit{1+2g});
WE denotes approaches based on word embeddings. Best value per column in bold.}
\label{tab:summary_results_retrieval}
\end{minipage}
\hfill
\begin{minipage}[t]{0.49\textwidth}
\centering
\resizebox{\linewidth}{!}{%
\begin{tabular}{lrrrrrr}
\toprule
\multirow{2}{*}{\textbf{Model}} & \multicolumn{3}{c}{\textbf{Verbatim (cit.)}} & \multicolumn{3}{c}{\textbf{Allusion (cf.)}} \\
\cmidrule(lr){2-4}\cmidrule(lr){5-7}
 & \textbf{P} & \textbf{R} & \textbf{F1} & \textbf{P} & \textbf{R} & \textbf{F1} \\
\midrule
\textbf{LogReg (lexical)} & 37.26 & 83.31 & 46.00 & 2.47 & 36.95 & 4.45 \\
\textbf{GBDT (lexical)} & 35.56 & \textbf{84.28} & 44.55 & 2.16 & 22.46 & 3.76 \\
\midrule
\textbf{mmBERT Base} & 75.35 & 75.44 & 73.67 & 8.54 & 44.66 & 12.67 \\
\textbf{mmBERT Small} & 71.94 & 77.01 & 71.86 & \textbf{8.90} & 37.08 & \textbf{12.79} \\
\textbf{XLM-R Large} & 75.44 & 81.07 & 75.65 & 6.68 & 53.23 & 10.87 \\
\textbf{XLM-R Base} & 68.54 & 74.57 & 68.71 & 6.77 & \textbf{55.83} & 10.91 \\
\textbf{mBERT Large} & 71.10 & 77.53 & 71.62 & 8.27 & 36.32 & 12.12 \\
\textbf{mBERT Base} & 68.73 & 74.64 & 69.10 & 8.07 & 41.61 & 12.03 \\
\textbf{LatinBERT} & \textbf{77.03} & 82.99 & \textbf{77.82} & 7.82 & 41.55 & 11.30 \\
\textbf{BERT-Romanian} & 62.10 & 70.52 & 63.29 & 7.26 & 42.14 & 11.33 \\
\textbf{PhilBERTa} & 61.75 & 72.98 & 63.65 & 2.80 & 45.25 & 5.04 \\
\textbf{LaBERTa} & 25.15 & 65.52 & 31.73 & 1.61 & 23.83 & 2.86 \\
\textbf{RoBERTa-Latin} & 0.81 & 26.48 & 1.53 & 1.02 & 11.54 & 1.74 \\
\bottomrule
\end{tabular}
}
\vspace{0.35em}
\captionof{table}{\textbf{Performance comparison of classification approaches.}
All values are $\times100$, macro-averaged over query segments and averaged over the 5 folds. Lexical approaches use sparse features; the remaining rows
are fine-tuned transformer encoders. Decision thresholds are tuned per class and model on the training set. Best value per column in bold.}
\label{tab:summary_results_classification}
\end{minipage}
\end{figure*}

\section{Experimental Setup}
\label{sec:framework}
\label{sec:methods}
\label{sec:experimental_setup}
\label{subsec:task_definition_main}

We use the dataset of Section~\ref{sec:dataset} to answer three questions: (i) how well do retrieval-only methods rank true source segments among many candidates; (ii) how well do classifiers separate verbatim references, allusions, and pairs labeled \textit{no match} on a fixed candidate set; and (iii) what precision--efficiency trade-off can a retrieve-and-rerank pipeline reach.

\paragraph{Task and pipeline.}
We define intertextuality detection as the identification of directional references between a \textit{query set} (segments from chronologically later works) and a \textit{source set} (segments from earlier works). A query segment can refer to one or more source segments, either as a verbatim reference or as an allusion; most query segments refer to none. Comparing 85k query against 91k source segments yields roughly 7.7 billion pairs, so exhaustive multiclass classification of every pair is computationally intractable. We therefore evaluate three method families: a retrieval-only approach that ranks candidates by similarity, a classification-only approach applied to a fixed candidate set, and a retrieve-and-rerank pipeline that classifies the top-$k$ candidates produced by a retriever as \textit{no match}, \textit{verbatim reference}, or \textit{allusion} (Figure~\ref{fig:two_stage_pipeline}).

\subsection{Retrieval Models}
\label{subsec:retrieval_models_main}
We compare three retriever families. The sparse lexical methods TF-IDF and BM25~\citep{robertson2009bm25} rank source segments by term-weighted similarity and have no learned parameters; their terms are either surface tokens or lemmas produced by the Classical Language Toolkit (CLTK; \citealp{johnson2021cltk}), taken as unigrams or unigrams+bigrams. Two word-embedding (WE) baselines reimplement prior work at the word level: following~\citet{DBLP:conf/naacl/BurnsBLCD21}, the mean cosine similarity of bigram Word2Vec~\citep{mikolov2013word2vec} vectors, and following~\citet{gong_augmented_2025}, position-aggregated LatinBERT~\citep{DBLP:journals/corr/abs-2009-10053} token embeddings. The dense retrievers are bi-encoders fine-tuned with Online Contrastive Loss~\citep{hadsell2006contrastive}; we evaluate the multilingual encoders BGE-M3~\citep{DBLP:journals/corr/abs-2402-03216}, E5-small/base/large~\citep{DBLP:journals/corr/abs-2402-05672}, and Granite-107m/278m~\citep{DBLP:journals/corr/abs-2502-20204}, as well as the Latin-specific SPhilBERTa~\citep{DBLP:conf/ranlp/Riemenschneider23}. Following the E5 recipe, query and candidate are encoded with distinct \texttt{Query:} and \texttt{Candidate:} prefixes. Each word-embedding baseline uses the tokenizer and vocabulary released with its original model, which matters most for LatinBERT. Grouped model descriptions and training details are in Appendix~\ref{app:retrieval_setup}.

\subsection{Classification Models}
\label{subsec:classification_models_main}
For classification we compare two families. As lexical reference points, we train a logistic regression (LogReg) and a histogram gradient-boosted decision tree (GBDT) on per-pair similarity features: TF-IDF cosine over lemma unigrams, lemma unigrams+bigrams, and character 3--4-grams, together with Jaccard overlap, raw-token overlap, and length features, all computed with the same CLTK pipeline as the sparse retrievers. Our main classifier is a cross-encoder that processes the query--candidate pair as a single concatenated input~\citep{DBLP:conf/emnlp/ReimersG19}, so self-attention can compare the two segments at the token level, which matters most for distinguishing allusions from verbatim references. We evaluate multilingual cross-encoders, namely mmBERT~\citep{DBLP:journals/corr/abs-2509-06888}, XLM-RoBERTa (XLM-R; \citealp{DBLP:conf/acl/ConneauKGCWGGOZ20}), ModernBERT (mBERT; \citealp{DBLP:conf/acl/WarnerCCWHTGBLA25}), and BERT-Romanian~\citep{DBLP:conf/emnlp/DumitrescuAP20}, together with the Latin-specific cross-encoders PhilBERTa and LaBERTa~\citep{DBLP:conf/acl/Riemenschneider23}, LatinBERT~\citep{DBLP:journals/corr/abs-2009-10053}, and RoBERTa-Latin~\citep{classcat2024robertalatin}. Table~\ref{tab:summary_results_classification} reports the small/base or base/large variants where applicable. All cross-encoders are fine-tuned for 4 epochs with batch size 32 and learning rate $2{\times}10^{-5}$; full feature definitions and training details are in Appendix~\ref{app:classification_setup}. The classifier is three-class throughout (\textit{no match}, \textit{verbatim reference}, \textit{allusion}); for retrieval-style evaluation (Recall@$k$ and overall reference-detection F1) the two positive classes are merged into a single positive label, while reference-type-specific metrics are obtained by filtering the same predictions.

\subsection{Metrics}
\label{subsec:metrics_main}
Because most query--source pairs are labeled \textit{no match}, accuracy is uninformative: a trivial all-negative predictor already scores near-perfectly. We therefore report per-class precision (P), recall (R), and F1 for verbatim references and allusions, and Recall@$k$ at $k \in \{10, 100, 1000\}$ for retrieval. Formal definitions and the global error-rate metrics (SMR, FPR, FNR) used in the appendix are given in Appendix~\ref{app:metrics}.

\paragraph{Aggregation.} Per-class P, R, and F1 are computed per query segment (one-vs-rest, at each model's tuned decision threshold), macro-averaged within each fold, and then averaged over the five folds (e.g., Table~\ref{tab:summary_results_classification}); Recall@$k$ is aggregated the same way.

\subsection{Data Splits}
\label{subsec:dataset_split_main}
The 1,490 references map onto 1,470 distinct query--source segment pairs, since 20 references share a pair with another reference. We use 5-fold cross-validation on these pairs, with all source segments of a given query assigned to the same fold; fold sizes therefore vary slightly.
For classification and retrieve-and-rerank evaluation, each fold contains a query document (citing query segments plus an equal number of in-domain distractors, doubling the total) and a source document (cited source segments plus an equal number of segments that are not cited, drawn from the same source works, again doubling the total), and every query segment is scored against every source segment in the fold. Per-fold sizes are reported in Table~\ref{tab:fold_sizes}: on average each fold spans $\sim$282 queries and $\sim$565 sources, yielding $\sim$159,000 candidate pairs of which $\sim$294 are ground-truth positives ($\sim$0.18\%) and the remainder serve as in-domain negatives. Retrieval-only evaluation instead ranks the held-out queries against the complete source corpus of 90,546 segments, which is identical across folds. Consequently, retrieval cutoffs such as Recall@1000 remain meaningful and need not saturate.

\begin{table}[t]
\centering
\small
\setlength{\tabcolsep}{3.5pt}
\resizebox{\columnwidth}{!}{%
\begin{tabular}{lrrrrrr}
\toprule
 & \multicolumn{2}{c}{\textbf{Queries}} & \multicolumn{2}{c}{\textbf{Sources}} & \multicolumn{2}{c}{\textbf{Pairs}} \\
\cmidrule(lr){2-3}\cmidrule(lr){4-5}\cmidrule(lr){6-7}
\textbf{Fold} & \textbf{Citing} & \textbf{+Distr.} & \textbf{Cited} & \textbf{+Distr.} & \textbf{Pos.} & \textbf{Neg.} \\
\midrule
0 & 142 & +142 & 269 & +269 & 277 & 152{,}515 \\
1 & 141 & +141 & 254 & +254 & 270 & 142{,}986 \\
2 & 141 & +141 & 319 & +319 & 331 & 179{,}585 \\
3 & 141 & +141 & 301 & +301 & 312 & 169{,}452 \\
4 & 141 & +141 & 269 & +269 & 280 & 151{,}436 \\
\midrule
Total & 706 & +706 & 1{,}412 & +1{,}412 & 1{,}470 & 795{,}974 \\
Mean & 141.2 & +141.2 & 282.4 & +282.4 & 294.0 & 159{,}195 \\
\bottomrule
\end{tabular}%
}
\caption{\textbf{Per-fold evaluation set sizes} for classification and retrieve-and-rerank. Each fold contains query segments (\emph{Citing}) that refer to at least one source segment (\emph{Cited}); both are padded 1:1 with distractors from the same works (\emph{+Distr.}). \emph{Pos.}\ are distinct annotated segment pairs and \emph{Neg.}\ all remaining query--source pairs.}
\label{tab:fold_sizes}
\end{table}

\subsection{Model Configuration and Ablation}
\label{subsec:model_configuration_main}
Each training fold is augmented with negative pairs at positive-to-negative ratios from $1{:}1$ to $1{:}10$. For the dense retriever and the cross-encoder we compare base model families, negative-sampling ratios, learning rates, and epochs over the same 5 folds; sparse and WE methods need no fine-tuning and use default configurations. Per-class thresholds are tuned on the training fold with a plateau heuristic selecting the largest threshold within 0.01 of the best F1, which favors precision. Full grids are in Appendices~\ref{app:retrieval_setup} and~\ref{app:classification_setup}, with the corresponding results in Appendices~\ref{app:retrieval_results} and~\ref{app:classification_results}. Code and pipelines are released as the Python package \texttt{locisimiles} (Appendix~\ref{app:software_toolkit}).

%% file: sections/07_results.tex
\section{Experimental Results}
\label{sec:experiments}

We address the three questions of Section~\ref{sec:experimental_setup} in turn: retrieval-only ranking (Section~\ref{subsec:ir_results}), classification on a fixed candidate set (Section~\ref{subsec:classification_results}), and the precision--efficiency trade-off of the retrieve-and-rerank pipeline (Section~\ref{subsec:rerank_results}).

\subsection{Information Retrieval Results}
\label{subsec:ir_results}

Lexical methods match or outperform dense retrievers (Table~\ref{tab:summary_results_retrieval}): BM25 with lemmatized unigrams achieves the best Recall@100 on both reference types. Among dense retrievers, BGE-M3 and E5-large lead and SPhilBERTa trails the E5 family; LatinBERT word embeddings beat every bi-encoder, while static word vectors fall far behind. Allusion recall stays far below verbatim throughout. Full results and ablations on negative sampling and hyperparameters are in Appendix~\ref{app:retrieval_results}.

Three factors help explain the advantage of lexical retrieval on allusions. First, lemmatization counters the fragmentation of Latin's rich morphology under subword tokenization, which benefits BM25. Second, bi-encoders compress each segment into a single pooled embedding, which can obscure a short local cue in otherwise divergent context; the \textit{nosce te} false negative in Section~\ref{subsec:qualitative_main} illustrates this, and per-pair correctness correlates with lexical overlap for verbatim references but not for allusions (Section~\ref{subsec:quantitative_breakdown}). Third, the benchmark retains a lexical component even among allusions, which share 1.65 lemmas on average; moreover, 275 of the 1,490 references were surfaced through n-gram candidate generation. As Section~\ref{subsec:classification_results} shows, this disadvantage is specific to pooled bi-encoder retrieval: cross-encoders compare both passages at the token level and outperform lexical classifiers on allusions.

\begin{figure*}[t]
    \centering
    \includegraphics[width=\linewidth]{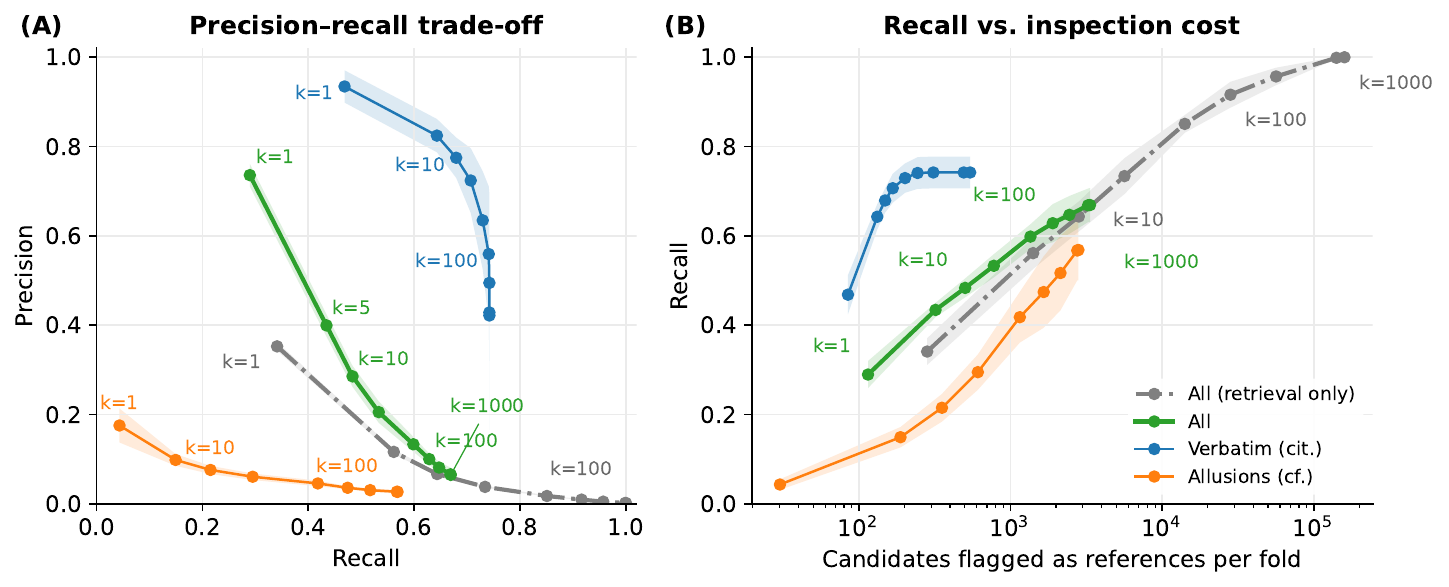}
    \vspace{-0.5cm}
    \caption{\textbf{Performance vs. efficiency trade-off in the retrieve-and-rerank pipeline.} The pipeline first retrieves the top-$k$ candidates with BM25 over lemma unigrams, then applies a cross-encoder classifier to assign the labels \textit{no match}, \textit{verbatim reference}, or \textit{allusion}. We compare this against a retrieval-only baseline where the top-$k$ candidates are treated as positive predictions. In (B), the number of candidates flagged as references per fold corresponds to the passages a philologist would need to inspect. Selected $k$ are labeled. Results are pooled per fold and averaged across 5 folds; shaded bands show $\pm1$ standard~deviation.}
    \label{fig:pipeline_tradeoff_short}
    \vspace{-0.25cm}
\end{figure*}

\subsection{Classification Results}
\label{subsec:classification_results}

Transformer encoders clearly outperform the lexical baselines (Table~\ref{tab:summary_results_classification}). LogReg and GBDT achieve high verbatim recall but low precision, and near-zero precision on allusions. LatinBERT and XLM-R Large lead on verbatim, mmBERT Small on allusions. Allusions remain far harder than verbatim references throughout, limited by low precision despite moderate recall. Full results and ablations on negative sampling and hyperparameters are in Appendix~\ref{app:classification_results}, with per-class breakdowns.

\subsection{Retrieve-and-Rerank Results}
\label{subsec:rerank_results}

Guided by the per-stage findings, we instantiate the retrieve-and-rerank pipeline of Section~\ref{sec:experimental_setup} with BM25 over lemma unigrams as the first stage, the strongest retriever on Recall@100 in Section~\ref{subsec:ir_results}, and the cross-encoder classifier as the reranker, which Section~\ref{subsec:classification_results} showed dominates the lexical baselines on precision. As a reference point we compare against a retrieval-only baseline that treats all top-$k$ candidates as detected references.

The pipeline produces far fewer false positives than retrieval-only (Figure~\ref{fig:pipeline_tradeoff_short}A). At $k=5$ it reaches F1 0.42 against 0.19, and at $k=100$ it still reaches 0.17 while retrieval-only falls to 0.02 (Table~\ref{tab:pipeline_comparison}). In practical terms, at $k=100$ the pipeline recovers 63\% of true intertextual references (185 out of 294) while narrowing the candidate set from 28,240 to 1,901 passages, a 93\% reduction in the space a philologist would need to inspect (Figure~\ref{fig:pipeline_tradeoff_short}B; Appendix~\ref{app:combined_results}).

%% file: sections/07b_error_analysis.tex
\section{Error Analysis}
\label{sec:error_analysis}

Complementing the aggregate results of Section~\ref{sec:experiments}, we now analyze where the XLM-R Large reranker succeeds and where it fails.

\subsection{Quantitative Breakdown}
\label{subsec:quantitative_breakdown}

The per-author-pair breakdown (Table~\ref{tab:xlmr_per_pair_class_fullgrid_mean_compact}) shows that classification quality depends largely on lexical overlap. For verbatim references the relationship is roughly monotonic: detection of high-overlap pairs is close to the ceiling, while the low-overlap Valerius pairs are the weakest. The same factor explains the verbatim--allusion gap, since allusions share far fewer lemmas. Within the allusion class, however, overlap no longer predicts performance: Valerius allusions outperform Jerome's despite lower overlap, suggesting that once overlap is scarce, the number of training pairs and source genre dominate. Valerius draws on well-attested poetic epic, whereas Jerome's sparse allusions to prose sources such as Cicero are effectively missed. The few near-perfect verbatim scores (F1 $=1.00$ for Jerome--Lucan, Jerome--Ovid, and Lactantius--Horace) rest on minimal support ($\leq 5$ pairs each) and should be read with caution. The breakdown separates two kinds of pair: those with enough shared lemmas to be matched, and those whose connection the surface text barely shows.

At the level of individual query--source pairs, the effect is strongly class-dependent (Figure~\ref{tab:overlap_correctness_corr}). For verbatim references, lexical overlap predicts correctness strongly (Spearman $\rho = 0.36$--$0.40$, $p<0.001$); for allusions the correlation collapses to near zero and is not significant; and for pairs labeled \textit{no match} it turns negative ($\rho \approx -0.09$), confirming that high overlap encourages spurious matches, the quantitative counterpart of the \textit{omnia timeo} false positive below. The weakly negative overall correlation is therefore an artifact of class balance: pairs labeled \textit{no match} dominate the pair grid, not evidence that overlap misleads in general.

\subsection{Qualitative Analysis}
\label{subsec:qualitative_main}

Two failure patterns recur. In a representative false positive, Jerome's generic \textit{omnia timeo} (``I fear everything'') overlaps with a Ciceronian passage, but his cue \textit{ut ait gentilis poeta} (``as the pagan poet says'') in fact points to Virgil's \textit{omnia tuta timens}, whose distinctive word \textit{tuta} is absent from the matched candidate: the model fires on a high-overlap distractor rather than the correct, lower-overlap target. In a representative false negative, the brief Socratic allusion \textit{nosce te} (``know yourself'') opens a Jerome passage that then shifts to an unrelated discussion of stylistic simplicity, and the pooled sentence embedding loses the local cue. Both failure modes point to token- or span-level matching for short, low-overlap references. Full text of both examples, source pairs, and philological discussion are in Appendix~\ref{subsec:qualitative_analysis}.

\begin{table}[t]
\centering
\scriptsize
\setlength{\tabcolsep}{3pt}
\resizebox{\columnwidth}{!}{%
\begin{tabular}{llcccccc}
\toprule
\multirow{2}{*}{\textbf{Query}} & \multirow{2}{*}{\textbf{Source}} & \multicolumn{3}{c}{\textbf{Verbatim (cit.)}} & \multicolumn{3}{c}{\textbf{Allusion (cf.)}} \\
\cmidrule(lr){3-5} \cmidrule(lr){6-8}
 & & \textbf{\#} & \textbf{Ov.} & \textbf{F1} & \textbf{\#} & \textbf{Ov.} & \textbf{F1} \\
\midrule
Jerome & Cicero & 91 & \ov{5.22} & \f{0.66} & 21 & \ov{3.14} & \f{0.04} \\
Jerome & Horace & 41 & \ov{6.80} & \f{0.98} & -- & -- & -- \\
Jerome & Lucan & 2 & \ov{5.00} & \f{1.00} & -- & -- & -- \\
Jerome & Ovid & 2 & \ov{6.00} & \f{1.00} & -- & -- & -- \\
Jerome & Virgil & 211 & \ov{6.95} & \f{0.89} & 38 & \ov{3.20} & \f{0.12} \\
Lactantius & Horace & 5 & \ov{12.40} & \f{1.00} & -- & -- & -- \\
Lactantius & Lucretius & 24 & \ov{13.21} & \f{0.99} & -- & -- & -- \\
Lactantius & Ovid & 23 & \ov{11.13} & \f{0.99} & -- & -- & -- \\
Lactantius & Propertius & 3 & \ov{8.00} & \f{1.00} & -- & -- & -- \\
Lactantius & Virgil & 78 & \ov{8.85} & \f{0.92} & -- & -- & -- \\
Valerius & Lucan & 65 & \ov{1.91} & \f{0.52} & 85 & \ov{1.33} & \f{0.32} \\
Valerius & Ovid & 55 & \ov{1.86} & \f{0.52} & 94 & \ov{1.16} & \f{0.29} \\
Valerius & Statius & 42 & \ov{2.26} & \f{0.49} & 78 & \ov{1.70} & \f{0.23} \\
Valerius & Virgil & 207 & \ov{2.13} & \f{0.53} & 305 & \ov{1.57} & \f{0.14} \\
\midrule
Total &  & 849 & \ov{5.11} & \f{0.76} & 621 & \ov{1.65} & \f{0.11} \\
\bottomrule
\end{tabular}
}
\caption{\textbf{Per-author-pair performance breakdown for XLM-R Large.} Support (\#), mean number of shared lemmas per pair (Ov.), and per-query macro F1 (averaged over 5 folds). Cells are shaded teal (darker = larger); ``--'' = no references of that type.}
\label{tab:xlmr_per_pair_class_fullgrid_mean_compact}
\vspace{-0.2cm}
\end{table}

%% file: sections/08_discussion.tex
\section{Discussion and Conclusion}
\label{sec:discussion}
\label{sec:conclusion}

In this work, we introduced \textit{Loci Similes}, a benchmark dataset for Latin intertextuality detection that evaluates whether language models capture semantic similarity beyond exact lexical matching. Our experiments show that while the evaluated methods recover many long verbatim references and thematic allusions, detecting subtle short parallels, i.e., references that share only two or three words, remains difficult.

BM25 with lemmatized unigrams achieved the best retrieval performance (Recall@100 of 77.6 for verbatim references) and LatinBERT the best classification results (verbatim F1 of 77.8, Tables~\ref{tab:summary_results_retrieval} and~\ref{tab:summary_results_classification}). Retrieving the top 100 candidates per query and reranking recovers 63\% of true intertextual references while reducing the candidate set by 93\%. The pipeline is a reference implementation rather than a production system (see Appendix~\ref{app:scalability} on scaling). Language models are therefore a promising avenue, but the central challenge, distinguishing meaningful reuse from coincidental lexical overlap, will require dedicated architectures.

\paragraph{Takeaways.} First, a lemma-level BM25 baseline outperforms every dense retriever we evaluate: Latin's rich morphology fragments under subword tokenizers, and CLTK lemmatization gives sparse methods an edge, motivating hybrid sparse--dense retrieval. The strongest neural retriever is not a pooled bi-encoder but token-level contextual embeddings, which nearly match the sparse retrievers on allusions. Second, neural cross-encoders show a clear advantage over lexical classifiers on allusions, with noticeably better precision on the subtler, low-overlap cases, while remaining competitive on verbatim references.

\begin{figure}[t]
\centering
\includegraphics[width=\columnwidth]{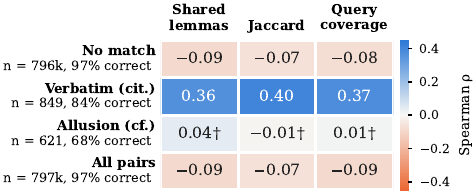}
\vspace{-0.3cm}
\caption{\textbf{Correlation between lexical overlap and per-pair correctness}, by gold class, for XLM-R Large on the full evaluation grid (Spearman $\rho$). All values significant at $p<0.001$ except those marked $\dagger$.}
\label{tab:overlap_correctness_corr}
\vspace{-0.1cm}
\end{figure}

\paragraph{Future Work.} Our error analysis suggests two directions: token-level matching (e.g., ColBERT~\citep{DBLP:conf/sigir/KhattabZ20}) with objectives rewarding semantic over lexical correspondence, since pooled embeddings lose faint signals of short allusions; and span- or clause-level matching, since signals are often local yet embedded in divergent context. Future work could also characterize how an author uses an earlier text (appropriation vs.\ polemical engagement). Extending the benchmark to further citing authors would reduce its dependence on a few source genres.

\section*{Limitations}
\label{sec:limitations}

\paragraph{Data Coverage.} Although our dataset comprises expert-verified positive pairs, it is not (and cannot be) exhaustive. The labeled dataset likely omits some valid references between the selected works. Consequently, it is possible that a few instances classified as ``false positives'' may represent genuine but undocumented intertextual references, potentially skewing the reported precision.

The 275 newly curated references were surfaced through n-gram candidate generation before philological verification. This procedure may favor references that retain some lexical overlap, so allusions without any shared lemma may be underrepresented within this subset. The remaining 1,215 references were adopted from existing scholarly datasets and did not pass through this candidate-selection step; the limitation therefore does not apply uniformly to the benchmark as a whole.

\paragraph{Operational Labels and Philological Ambiguity.} Defining what constitutes a ``reference'' in classical philology remains a substantial methodological challenge. The boundaries between literal citation, subtle allusion, and general thematic resonance are fluid and subject to differing scholarly definitions. For a reproducible benchmark, however, annotators do not have to settle this question. Once an intertextual reference has been established, its benchmark type is assigned using the operational lemma-, proximity-, and punctuation-based criterion described in Section~\ref{subsec:annotation}. While the taxonomy in Figure~\ref{fig:intertext_taxonomy} illustrates broader philological interpretations, its subtypes do not alter the operational assignment. The resulting labels should therefore be understood as reproducible analytical categories rather than exhaustive interpretations of each intertext. Accordingly, the low F1 scores for allusions indicate that models struggle to predict this operational category; they do not, by themselves, demonstrate inconsistency or unreliability in the labels.

\section*{AI Usage Statement}

Language model-based AI tools (Codex and GitHub Copilot) were used as coding assistants during implementation. Additionally, AI-based tools were used for grammar and spell checking. No AI-generated text appears in the final manuscript.

\section*{Acknowledgements}

This work was supported by the Deutsche Forschungsgemeinschaft (DFG, German Research Foundation), project no.\ 382880410, \textit{Quoting as a narrative strategy: A digital-hermeneutic analysis of intertextual references in Christian Late Antiquity}. We further thank the philologists whose expert annotation and verification made the benchmark possible, and the authors of the datasets we build on for making their resources publicly available.

%% file: sections/appendix/09_sources.tex
\section{Corpus Sources}
\label{app:sources}

The corpus texts were aggregated from three digital repositories: \textit{Corpus Corporum}\footnote{\href{https://mlat.uzh.ch/}{https://mlat.uzh.ch/}}, the \textit{Tesserae Project}\footnote{\href{https://github.com/tesserae/tesserae}{https://github.com/tesserae/tesserae}}, and the \textit{OpenGreekandLatin Project}. Table~\ref{tab:sources} lists the editions used; Figure~\ref{fig:dataset_examples} illustrates representative intertextual references from the ground truth.

\begin{table}[htbp]
\centering
\resizebox{\linewidth}{!}{%
\begin{tabular}{llcl}
\toprule
\textbf{Author} & \textbf{Work} & \textbf{Source} & \textbf{Edition} \\
\midrule
Virgil& Aeneid & CC & Greenough (1900) \\
Virgil& Georgics & CC & Greenough (1900) \\
Virgil& Eclogues & CC & Greenough (1900) \\
\midrule
Ovid & Amores & CC & Ehwald (1907) \\
Ovid & Ars Amatoria & CC & Ehwald (1907) \\
Ovid & Ex Ponto & CC & Wheeler (1939) \\
Ovid & Fasti & CC & Frazer (1933) \\
Ovid & Heroides & CC & Ehwald (1907) \\
Ovid & Ibis & CC & Merkel/Ehwald (1889) \\
Ovid & Medicamina & CC & Ehwald (1907) \\
Ovid & Metamorphoses & CC & Magnus (1892) \\
Ovid & Remedia Amoris & CC & Ehwald (1907) \\
Ovid & Tristia & CC & Wheeler (1939) \\
\midrule
Martial & Epigrammata & CC & Heraeus (1925) \\
Lucretius & De Rerum Natura & CC & Martin (1934) \\
Lucan & Pharsalia & CC & Weise (1835) \\
\midrule
Horace & Carmen Saeculare & CC & Shorey (1898) \\
Horace & Carmina & CC & Shorey (1919) \\
Horace & Ars Poetica & CC & Smart (1836) \\
Horace & Epistulae & CC & Fairclough (1929) \\
Horace & Epodes & CC & Vollmer (1912) \\
Horace & Saturae & CC & Smart (1836) \\
\midrule
Catullus & Carmina & Tess & Merrill \\
Propertius & Elegiae & Tess & Mueller (1898) \\
Tibullus & Elegiae & Tess & Postgate (1915) \\
Cicero & \textit{Opera Omnia} & Tess & \textit{Varia}\\
\midrule
Jerome& Epistulae & OGL & Hilberg (1910) \\
Jerome& \textit{Varia} & CC & Patrologia Latina (1845) \\
\midrule
Valerius Flaccus & Argonautica & CC & Kramer (1913) \\
Statius & Thebais & CC & Mozley (1928) \\
\bottomrule
\end{tabular}%
}
\caption{\textbf{Corpus data sources.} Abbreviations: CC (Corpus Corporum), Tess (Tesserae Project), and OGL (OpenGreekandLatin Project).}
\label{tab:sources}
\end{table}

%% file: sections/appendix/10_setup.tex
\section{Evaluation Metrics}
\label{app:metrics}

The classification model and the full retrieve-and-rerank pipeline are evaluated on the same directional comparison task between \textit{query} and \textit{source} documents, using 5-fold cross-validation over the 1,470 distinct positive segment pairs.

Unlike standard information retrieval tasks that focus on ranking top-$k$ candidates, our objective is to classify the full set of possible references between the query and source documents. Because many query segments have no true positive match in the source document, evaluation must reward the correct rejection of pairs labeled \textit{no match} as well as the recovery of attested references. We therefore define $N = \text{TP} + \text{FP} + \text{FN} + \text{TN}$ as the total number of text-segment pairs and use the following error-based metrics, which better match the practical constraints of philological workflows.

\begin{itemize}[leftmargin=10pt, itemsep=1pt,topsep=10pt]
    \item \textbf{Segment-Misclassification Rate (SMR):} defined as the fraction of all query-source pairs that were misclassified. This serves as a global error rate.
    Values range from $0$ (perfect retrieval) to $1$ (complete failure).
    $$\text{SMR} = \frac{\text{FP} + \text{FN}}{N}$$

    \item \textbf{Global False-Positive Rate (FPR):} defined as the share of the total dataset incorrectly predicted as references. A high FPR indicates a system prone to ``over-generating'' candidate references.
    $$\text{FPR} = \frac{\text{FP}}{N}$$

    \item \textbf{Global False-Negative Rate (FNR):} defined as the share of the total dataset that contains true references missed by the system. A high FNR indicates that genuine intertextual references remain undiscovered.
    $$\text{FNR} = \frac{\text{FN}}{N}$$
    
\end{itemize}

Together, these metrics decompose the total error (SMR) into false-positive and false-negative components, making it clear whether a model tends to over-generate candidate references or miss genuine references. We calculate each metric per query segment and report the mean across queries.

\section{Data Split and Evaluation Setup}
\label{app:data_split}

All retrieval, classification, and retrieve-and-rerank experiments use the same 5-fold split over the 1,470 distinct positive segment pairs. The split is stratified at the query level: all positive targets of a given query segment are assigned to the same fold, so no query appears in both training and held-out evaluation. Fold sizes therefore vary slightly, because individual queries can cite different numbers of source segments, even though the number of query segments per fold is approximately balanced.

\paragraph{Document-level, fully pairwise evaluation.} For classification and retrieve-and-rerank evaluation, each fold uses a reduced pairwise document pool. The \emph{query document} contains held-out query segments plus an equal number of query-side distractors drawn from the same work but without annotated references; the \emph{source document} contains the ground-truth source segments plus an equal number of source-side distractors drawn from the cited works. We then score \emph{every} query segment against \emph{every} source segment in that fold. The evaluation is therefore fully pairwise rather than restricted to gold-aligned pairs, forcing models to distinguish true references from in-domain distractors written by the same authors and in similar registers. Retrieval-only experiments use the same held-out query split but rank each query against the complete source corpus of 90,546 segments, which is identical in every fold.

\paragraph{Per-fold sizes.} Per-fold sizes (queries, sources, ground-truth pairs, and total pairwise comparisons) are reported in Table~\ref{tab:fold_sizes} in Section~\ref{subsec:dataset_split_main}.

The reduced per-fold query and source documents are reused across classification models and retrieve-and-rerank runs; all retrieval-only models use the separate full-corpus pool described above. Each fold then requires roughly $1.6\times10^{5}$ pairwise scores, of which only $\sim 0.18\%$ are true positives; the global error-rate metrics in Appendix~\ref{app:metrics} make this imbalance visible.

\section{Retrieval Setup}
\label{app:retrieval_setup}

This appendix covers the retrieval models, ablations, negative sampling, and implementation; results are in Appendix~\ref{app:retrieval_results}.

\subsection{Model Families}
\label{app:retrieval_model_families}

We group the retrieval methods by the kind of signal they exploit, from sparse lexical overlap to fine-tuned dense bi-encoders.

\paragraph{Sparse retrievers.} We evaluate TF-IDF and BM25 over both surface tokens and CLTK lemmas, using either unigrams alone or unigrams plus bigrams. For each query segment, source segments are ranked by sparse vector similarity (TF-IDF) or BM25 score, without learned parameters. TF-IDF uses a custom analyzer over the CLTK-tokenized input with $\texttt{sublinear\_tf}=\text{True}$, $\texttt{max\_features}=50\text{k}$, $\texttt{min\_df}=1$, and $\texttt{max\_df}=1.0$, fit on the training fold only.

\paragraph{Word-embedding retrievers.} We reimplement two prior approaches that operate on word-level rather than sentence-level embeddings. Following~\citet{DBLP:conf/naacl/BurnsBLCD21}, we score segment pairs by the mean cosine similarity of their bigram Word2Vec vectors, testing a Bamman--Burns lemma Word2Vec checkpoint\footnote{\texttt{qcl\_bamman\_lemma\_300}} and a LiLa fastText skip-gram checkpoint.\footnote{\texttt{lila\_fasttext\_skip\_win5\_min5}} Following~\citet{gong_augmented_2025}, we use position-aggregated contextual LatinBERT\footnote{\texttt{ashleygong03/bamman-burns-latin-bert}, tokenized with the subword encoder released with the original model; a WordPiece tokenizer over its converted vocabulary instead maps about 60\% of Latin words to \texttt{[UNK]}. The same encoder is used for the fine-tuned LatinBERT cross-encoder of Table~\ref{tab:summary_results_classification}. CLTK lemmas are mapped to the u/i orthography of the static vocabularies before lookup.} token embeddings as the signal.

\paragraph{Dense retrievers.} For sentence-level embedding, we evaluate the multilingual E5 family (Small, Base, and Large) \cite{DBLP:journals/corr/abs-2402-05672}, the Granite embedding models (107m and 278m) \cite{DBLP:journals/corr/abs-2502-20204}, and BGE-M3 \cite{DBLP:journals/corr/abs-2402-03216}. We also include SPhilBERTa \cite{DBLP:conf/ranlp/Riemenschneider23}, a sentence-transformer model for Ancient Greek, Latin, and English built on the PhilBERTa encoder \cite{DBLP:conf/acl/Riemenschneider23}. All dense retrievers are used as bi-encoders that embed queries and sources independently and rank candidates by cosine similarity.

\subsection{Ablation Design}
\label{app:retrieval_ablations}

We evaluate model choice and training regime for the retrieval stage independently before combining the selected components in the retrieve-and-rerank pipeline.

\begin{itemize}
    \item \textbf{Base-model ablations.} We compare sparse lexical retrievers, word-embedding baselines, and fine-tuned bi-encoders to isolate the effect of lexical matching, word-level semantics, and sentence-level dense retrieval.
    \item \textbf{Negative-sampling ablations.} For the learned bi-encoder, we compare multiple strategies for constructing negatives and vary the positive-to-negative training ratio over the grid $1{:}1, 1{:}2, \ldots, 1{:}10$ (ten ratios) to measure sensitivity to class imbalance.
    \item \textbf{Hyperparameter ablations.} For the fine-tuned bi-encoder, we vary the learning rate $\in\{1{\times}10^{-5}, 2{\times}10^{-5}, 3{\times}10^{-5}\}$ and number of epochs $\in\{1,2,4,6\}$ (12 configurations $\times$ 5 folds $=$ 60 runs) to identify stable operating regions rather than relying on a single default setting.
    \item \textbf{Pipeline-depth ablations.} For the full retrieve-and-rerank system, we vary retrieval depth over 15 values, from $k=1$ to $k=1000$ (1, 5, 10, 20, 50, and then 100--1000 in steps of 100), to measure the trade-off between recall, false positives, and the number of candidate pairs inspected or reranked. A retrieval-only baseline treats the top-$k$ candidates as positive predictions; the reranker is applied only to this pool (results in Appendix~\ref{app:combined_results}).
\end{itemize}

\subsection{Negative Sampling}
\label{app:retrieval_negative_sampling}

The quality of the learned bi-encoder depends on the negative examples used during training. We evaluated four negative sampling strategies:

\begin{itemize}
    \item \textbf{Random pairs ($\langle \text{random}, \text{random} \rangle$):} Pairs are formed by selecting two completely disjoint segments at random from the corpus.
    \item \textbf{Random negatives ($\langle \text{query}, \text{random} \rangle$):} For each positive query, we sample a negative candidate uniformly at random from the remaining corpus.
    \item \textbf{Hard negatives ($\langle \text{query}, \text{similar} \rangle$):} A pre-trained embedding model identifies ``hard negatives'': candidates that are semantically similar to the query but are not intertextual references.
    \item \textbf{Mixed negatives ($\langle \text{query}, \text{mix} \rangle$):} Uniformly sampled negatives and semantically similar hard negatives are combined for the same query.
\end{itemize}

\subsection{Implementation Details}
\label{app:retrieval_training}

We evaluate the sparse retrievers (TF-IDF, BM25) over both surface tokens and CLTK lemmas, with unigrams alone and unigrams plus bigrams. All vectorizers share the CLTK tokenization and lemmatization pipeline and are fit on the training fold only.

For the dense retriever, we employ sentence-transformer bi-encoders that generate embeddings for query and source segments independently. During inference, the bi-encoder computes cosine similarity between the query embedding and every source embedding in the index, and candidates are ranked by this score. To adapt the retriever to the task, we fine-tune the model using Online Contrastive Loss, which pulls linked pairs together and pushes unrelated pairs apart.

We fine-tuned the bi-encoder for 4 epochs (batch size 32, learning rate $2\times10^{-5}$, AdamW with $\beta_1=0.9$, $\beta_2=0.999$, $\epsilon=10^{-8}$, $\texttt{max\_grad\_norm}=1.0$, weight decay 0.01) using a linear learning-rate scheduler with 10\% warmup and, following the E5 recipe~\cite{DBLP:journals/corr/abs-2402-05672}, prepended ``Query: '' to query segments and ``Candidate: '' to source segments before encoding. Evaluation during training uses batch size 8 and is run every 50 optimizer steps; we use a fixed seed of 42 across folds. The bigram Word2Vec retriever uses a public checkpoint trained on the Bamman--Burns Latin corpus.

\section{Classification Setup}
\label{app:classification_setup}

The multiclass reranking setup covers model families, ablation design, sampling strategies, and implementation details. The reranker assigns one of three labels, $0 = \textit{no match}$, $1 = \textit{cit.}$, or $2 = \textit{cf.}$, to each (query, candidate) pair. Classification results are reported in Appendix~\ref{app:classification_results}.

\subsection{Model Families}
\label{app:classification_model_families}

We group the classifiers by the kind of signal they exploit.

\paragraph{Sparse classifiers.} As lexical reference points for reranking, we train a logistic regression (LogReg) and a histogram gradient-boosted decision tree (GBDT) on pairwise similarity features computed for each (query, candidate) pair. These models test how far explicit lexical evidence alone can separate \textit{no match}, \textit{verbatim reference}, and \textit{allusion}.

\paragraph{Neural cross-encoders.} For reranking, we evaluate multilingual and domain-adapted transformer encoders as multiclass cross-encoders. The multilingual baselines include XLM-R (Base and Large) \cite{DBLP:conf/acl/ConneauKGCWGGOZ20}, ModernBERT (Base and Large) \cite{DBLP:conf/acl/WarnerCCWHTGBLA25}, and mmBERT (Small and Base) \cite{DBLP:journals/corr/abs-2509-06888}. To measure the impact of domain adaptation, we evaluate PhilBERTa and LaBERTa \citep{DBLP:conf/acl/Riemenschneider23}, LatinBERT \citep{DBLP:journals/corr/abs-2009-10053}, and RoBERTa-Latin \citep{classcat2024robertalatin}, which are pre-trained on Latin corpora. We also include BERT-Romanian \cite{DBLP:conf/emnlp/DumitrescuAP20} to test cross-lingual transfer from a related Romance language. Unlike the bi-encoder retrievers, cross-encoders process the query and candidate jointly and predict one of the three relation labels directly.

\subsection{Ablation Design}
\label{app:classification_ablations}

We evaluate model choice and training regime for the reranking stage independently before combining the selected components in the retrieve-and-rerank pipeline.

\begin{itemize}
    \item \textbf{Base-model ablations.} We compare sparse lexical classifiers against multilingual and domain-adapted transformer cross-encoders.
    \item \textbf{Negative-sampling ablations.} We compare multiple strategies for constructing negatives and vary the positive-to-negative training ratio over the grid $1{:}1, 1{:}2, \ldots, 1{:}10$ (ten ratios) to measure sensitivity to class imbalance. Pair labels are looked up in a precomputed (query, corpus) table; when a pair carries both a \textit{cit.}\ and a \textit{cf.}\ annotation, \textit{cit.}\ takes precedence.
    \item \textbf{Three-way class-sampling ablations.} In a follow-up experiment for the 3-class classifier, we vary the target sampling proportions of the three training labels, $0 = \textit{no match}$, $1 = \textit{cit.}$, and $2 = \textit{cf.}$, using a target-class-probability sampler. The default rerun uses a mild fixed exposure of $\{0{:}0.80, 1{:}0.10, 2{:}0.10\}$; the ablation also tests $\{0.50, 0.25, 0.25\}$, $\{0.60, 0.20, 0.20\}$, $\{0.33, 0.33, 0.33\}$, and a ``natural'' setting without resampling. This tests how strongly the reranker depends on the relative frequency of negatives, verbatim references, and allusions during training.
\end{itemize}

\subsection{Negative Sampling and Class Sampling}
\label{app:classification_negative_sampling}

The quality of the multiclass reranker depends on the negative examples and class-mixing strategy used during training. We evaluated four negative sampling strategies:

\begin{itemize}
    \item \textbf{Random pairs ($\langle \text{random}, \text{random} \rangle$):} Pairs are formed by selecting two completely disjoint segments at random from the corpus.
    \item \textbf{Random negatives ($\langle \text{query}, \text{random} \rangle$):} For each positive query, we sample a negative candidate uniformly at random from the remaining corpus.
    \item \textbf{Hard negatives ($\langle \text{query}, \text{similar} \rangle$):} A pre-trained embedding model identifies ``hard negatives'': candidates that are semantically similar to the query but are not intertextual references.
    \item \textbf{Mixed negatives ($\langle \text{query}, \text{mix} \rangle$):} Uniformly sampled negatives and semantically similar hard negatives are combined for the same query.
\end{itemize}

In addition to choosing how negatives are constructed, the multiclass reranker requires a decision about how often each of the three labels is shown during training. After remapping the labels to $0 = \textit{no match}$, $1 = \textit{cit.}$, and $2 = \textit{cf.}$, training batches are drawn with a target-class-probability sampler rather than relying only on the raw empirical class frequencies. The default setting uses class probabilities $\{0{:}0.80, 1{:}0.10, 2{:}0.10\}$, and the ablation varies these three-way proportions to measure the effect of emphasizing negatives versus the two positive reference types during training.

\subsection{Implementation Details}
\label{app:classification_training}

The sparse classifiers (logistic regression and GBDT) use pairwise features for each (query, candidate) pair: TF-IDF cosine similarities over lemma unigrams, lemma unigrams+bigrams, and character 3--4-grams, plus Jaccard overlap on lemmas, raw token overlap, and length features. All vectorizers share the retrievers' CLTK tokenization and lemmatization pipeline and are fit on the training fold only.

For the cross-encoder, we concatenate query and candidate, separated by the model's special tokens (Figure~\ref{fig:two_stage_pipeline}), and truncate each to $50\%$ of the token budget to prevent either from dominating.
$$\texttt{<s>} \; \textit{Legimus ...} \; \texttt{</s>}\texttt{</s>} \; \textit{excudent...} \; \texttt{</s>}$$

We fine-tuned the cross-encoder for 4 epochs (per-GPU batch size 32 for base models and 16 for large models, with gradient accumulation chosen to keep an effective batch size of 64 across GPUs, learning rate $2\times10^{-5}$, AdamW) to predict the multiclass relation label for each query-candidate pair and minimize cross-entropy against the ground-truth labels. RoBERTa-family encoders use an explicit separator fix to insert the correct \texttt{</s></s>} sequence between query and candidate.

In the 3-class classifier follow-up, the cross-encoder uses a target-class-probability sampler during training. This sampler is implemented as a \texttt{WeightedRandomSampler} that is re-drawn each epoch, so that the empirical class exposure matches the configured target within each fold. We do not additionally re-weight the cross-entropy loss, so the sampler is the only mechanism used for the sampling-proportion ablation described above.

For reporting, per-class decision thresholds are tuned on the training fold using a \textit{plateau-high} heuristic with tolerance $0.01$: starting from the highest threshold, we walk down the precision--recall curve and pick the largest threshold at the top of the F1 plateau, independently for the \textit{cit.}\ and \textit{cf.}\ classes. The remaining mass is assigned to \textit{no match}. These tuned thresholds are then frozen and applied to the held-out fold.

\section{Retrieve-and-Rerank Setup}
\label{app:combined_setup}

This experiment evaluates the full retrieve-and-rerank system on the fold-level evaluation sets described in Appendix~\ref{app:data_split}. Unlike retrieval-only evaluation, this experiment uses the reduced fold-level source pool. For each held-out fold, the retriever first scores every query segment against the source document and returns the top-$k$ candidates; the reranker then classifies only this candidate pool. All remaining query--source pairs are treated as negative predictions. This preserves the fully pairwise document-level evaluation while measuring how much of the exhaustive comparison grid can be avoided through candidate generation.

The final retrieve-and-rerank configuration uses BM25 with lemma unigrams (the strongest sparse retriever; see Appendix~\ref{app:retrieval_results}) as the candidate generator. The reranking stage uses the fine-tuned XLM-R Large cross-encoder; the pipeline was not re-run with LatinBERT, whose verbatim F1 is slightly higher. For each candidate pair, the reranker predicts $0 = \textit{no match}$, $1 = \textit{cit.}$, or $2 = \textit{cf.}$; the two positive classes are merged when reporting reference-detection performance, and reference-type-specific metrics are computed by filtering the same cached predictions.

We compare this retrieve-and-rerank system against a retrieval-only baseline at the same retrieval depths. The retrieval-only baseline assigns all top-$k$ candidates to the positive class and all other pairs to \textit{no match}; the retrieve-and-rerank pipeline instead lets the classifier decide which retrieved candidates remain positive. Retrieval depth is swept over 15 values, from $k=1$ to $k=1000$ (1, 5, 10, 20, 50, and then 100--1000 in steps of 100); the reported retrieve-and-rerank results in Appendix~\ref{app:combined_results} focus on representative depths from this grid.

%% file: sections/appendix/11_results.tex
\section{Retrieval Results}
\label{app:retrieval_results}

\begin{figure*}[tp]
    \centering
    \includegraphics[width=\linewidth]{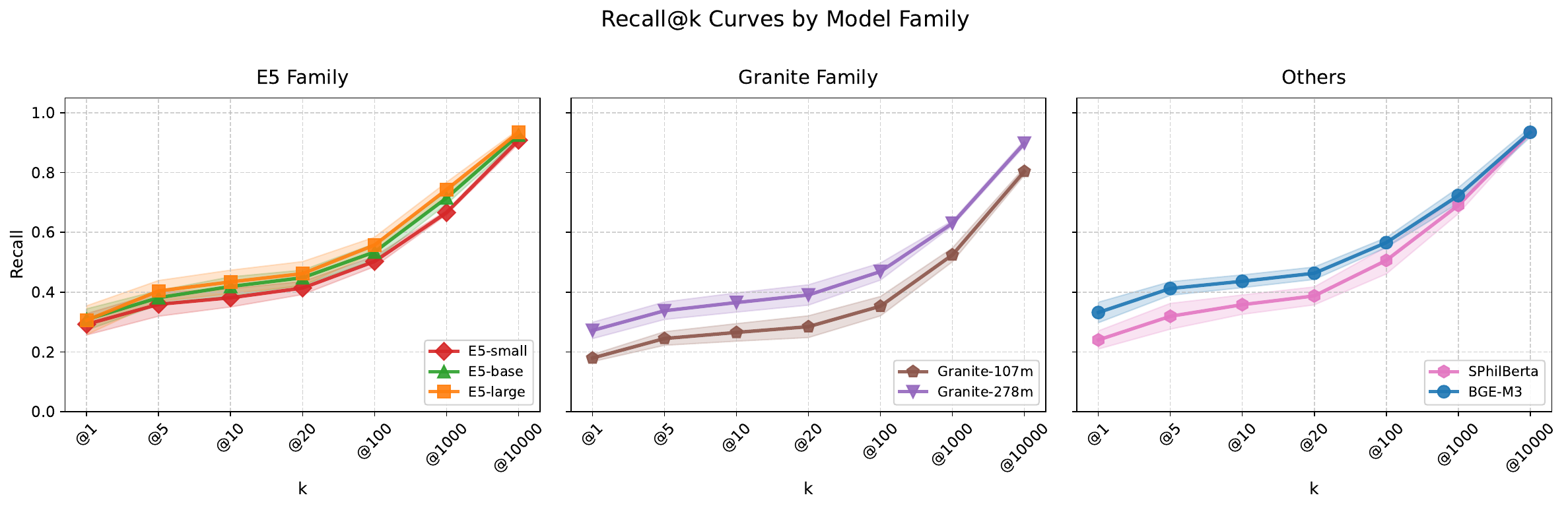}
    \caption{\textbf{Recall@$k$ comparison across embedding model families.} Performance of E5, Granite, BGE-M3, and the domain-specific SPhilBERTa on the Latin intertextuality retrieval task. We observe that larger models show better performance, with multilingual E5-large achieving the largest scores overall. Error bands show $\pm$1 std across 5-fold cross-validation.}
    \label{fig:embedding_model_comparison}
\end{figure*}

\begin{figure*}[tp]
    \centering
    \includegraphics[width=\linewidth]{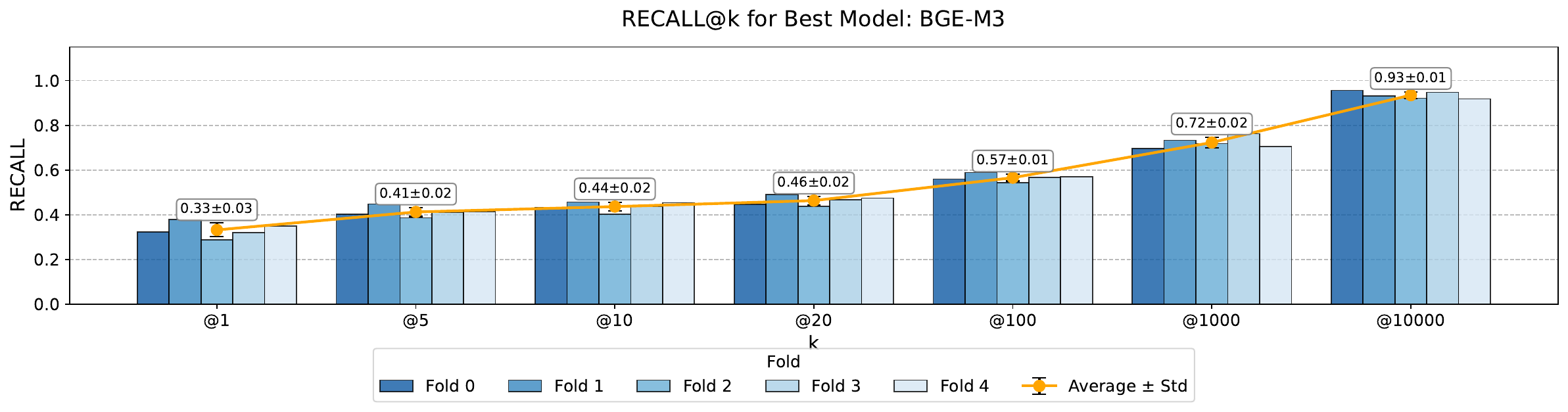}
    \caption{\textbf{Recall@$k$ performance of the best embedding model across individual folds.} We show recall at varying cutoff values ($k \in \{1, 5, 10, 20, 100, 1000, 10000\}$) for the top-performing model on the Latin intertextuality retrieval task. Each bar represents a single fold from 5-fold cross-validation, with the orange line indicating mean $\pm$ std. Performance improves substantially with larger $k$, approaching near-perfect recall at $k=10000$.}
    \label{fig:embedding_best_model_recall}
\end{figure*}

\begin{figure*}[tp]
    \centering
    \includegraphics[width=\linewidth]{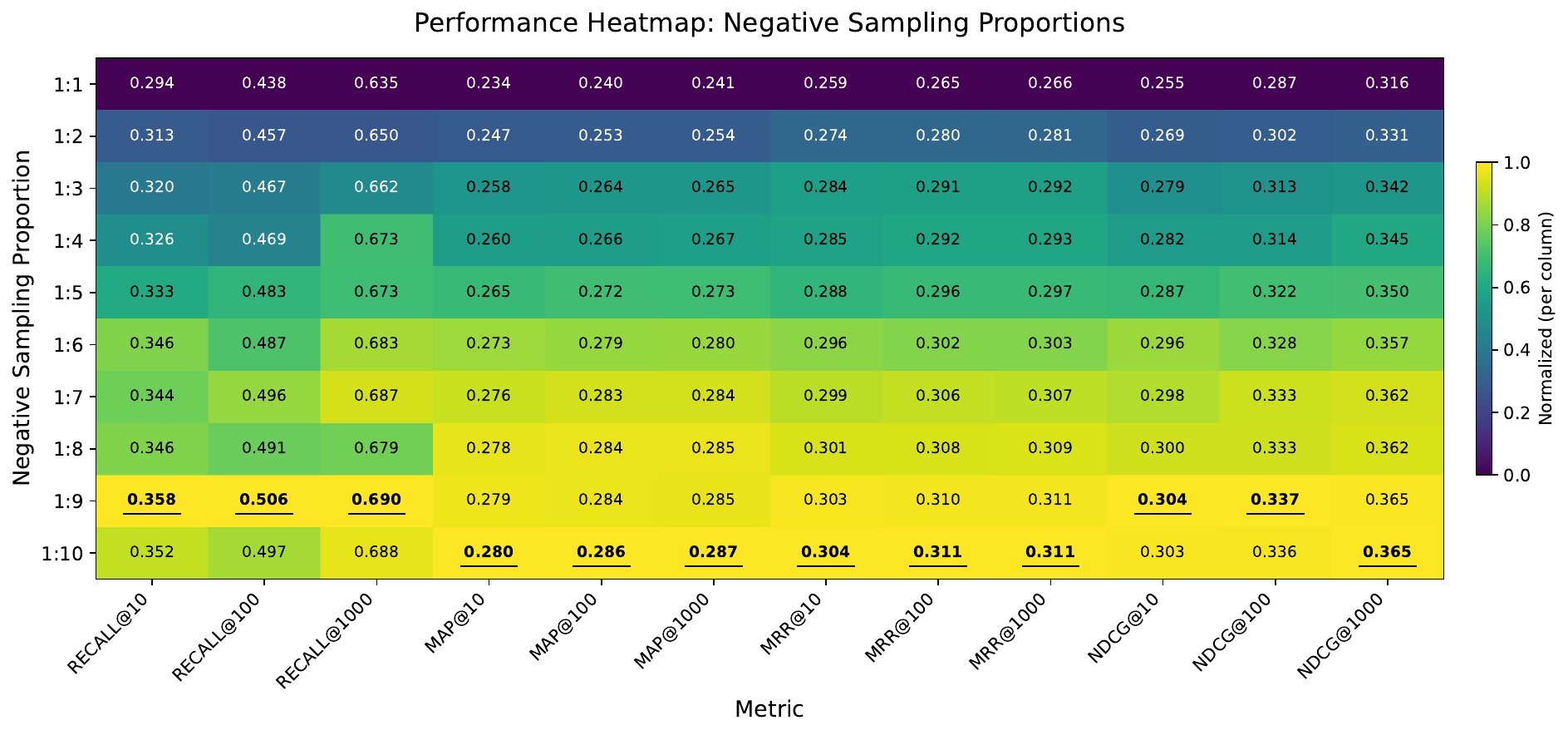}
    \caption{\textbf{Impact of training data imbalance on retrieval.} Heatmap showing recall, MAP, MRR, and NDCG scores at different cutoff values (@10, @100, @1000) across negative sampling proportions (1:1 to 1:10). Values are averaged over 5 cross-validation folds. Colors are normalized per column to highlight relative performance differences. Best values per metric are shown in bold with an orange underline. Larger ratios of negative samples generally improve retrieval performance, with optimal results often achieved at ratios of 1:5 or larger.}
    \label{fig:ir_neg_proportion}
\end{figure*}

\begin{figure*}[tp]
    \centering
    \includegraphics[width=0.85\linewidth]{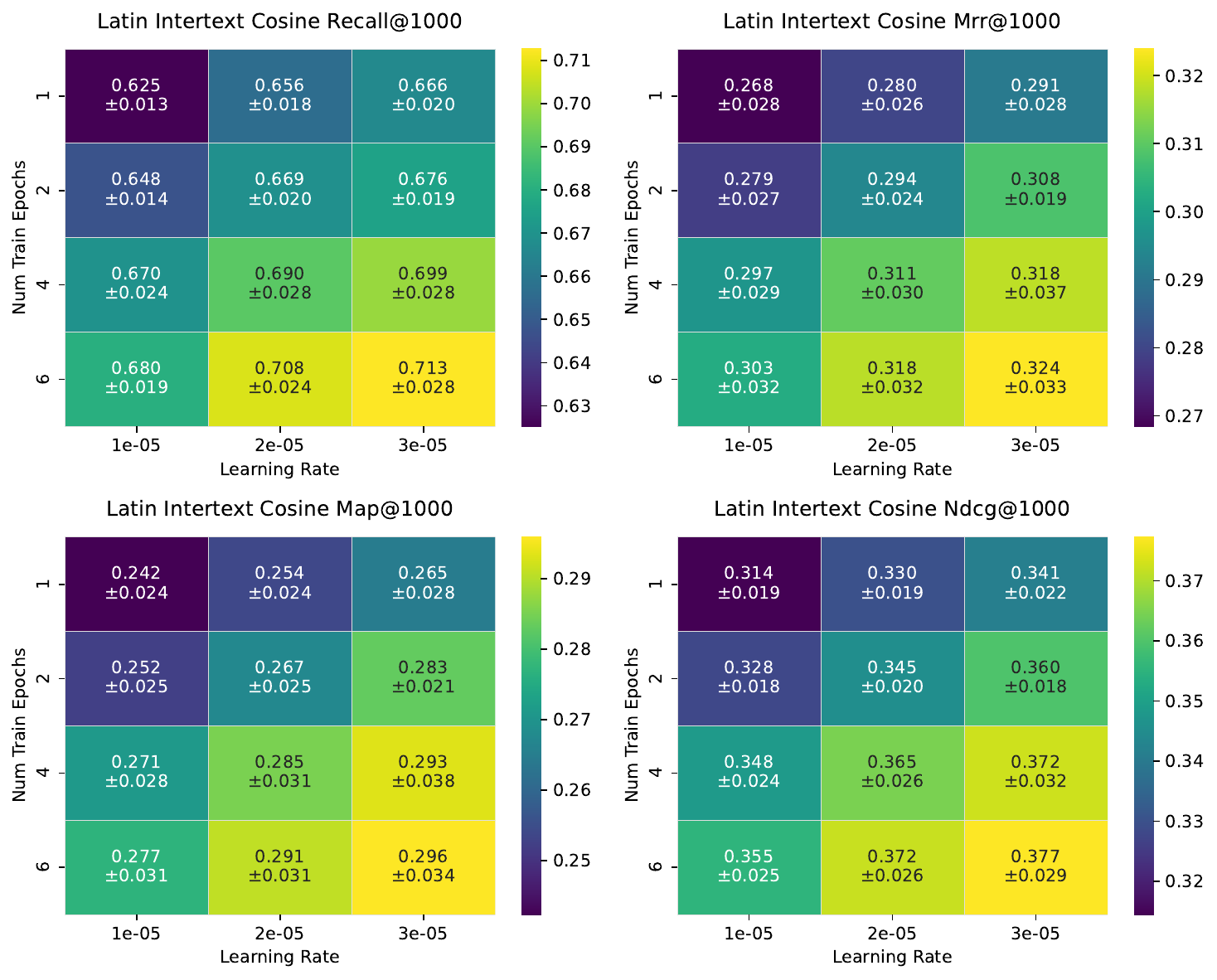}
    \caption{\textbf{Hyperparameter sensitivity for the dense retriever.} We visualize retrieval metrics (recall, MRR, MAP, and NDCG at $k=1000$) across varying learning rates and training epochs. Results are averaged across 5 folds, where annotated values represent the mean score $\pm$ standard deviation.}
    \label{fig:ir_param_tuning}
\end{figure*}

We evaluate candidate retrieval with standard information-retrieval metrics, focusing on whether the first stage preserves enough true references for downstream reranking. 
Table~\ref{tab:ir_model_results} and Figure~\ref{fig:embedding_model_comparison} show that the larger multilingual embedding models have the strongest recall among the dense retrievers. Table~\ref{tab:ir_baseline_results} reports the lexical and word-embedding retrieval baselines. Figure~\ref{fig:embedding_best_model_recall} gives the fold-level behavior of the best-performing model, while Figure~\ref{fig:ir_neg_proportion} and Figure~\ref{fig:ir_param_tuning} summarize the negative-ratio and hyperparameter ablations.

\begin{figure*}[tp]
    \centering
    \includegraphics[width=\linewidth]{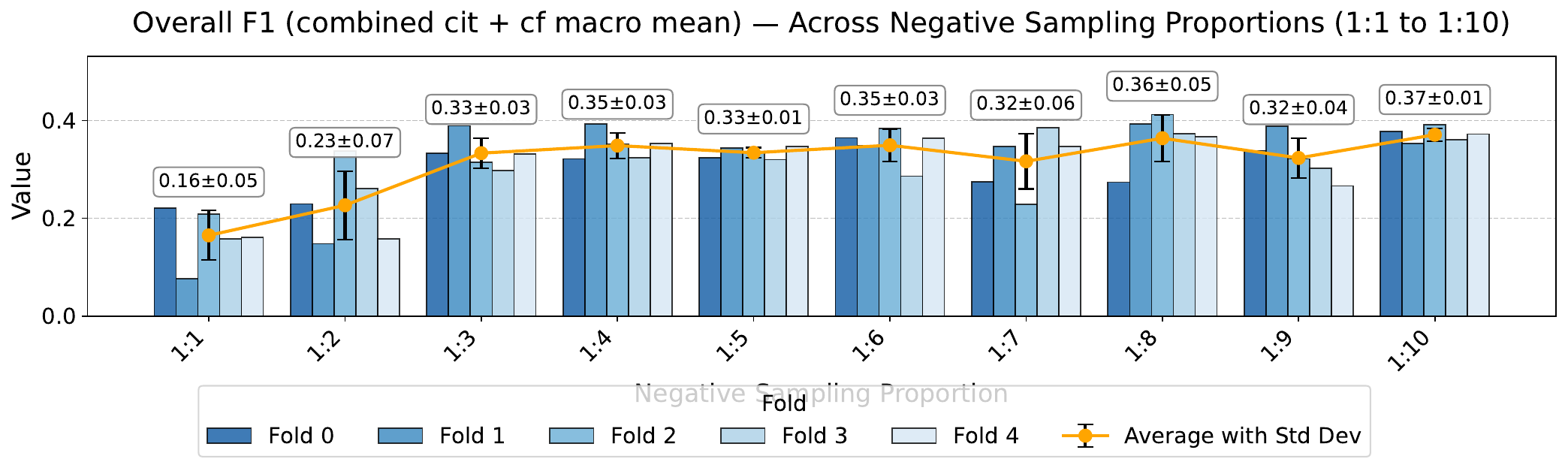}
    \caption{\textbf{Impact of training data imbalance on classification performance.} The plot illustrates the overall F1 (macro mean over \textit{cit.} and \textit{cf.}) on the evaluation set as the positive-to-negative training ratio is increased from $1{:}1$ to $1{:}10$. Shaded areas represent the standard deviation across the 5 cross-validation folds.}
    \label{fig:neg_sample_proportion}
\end{figure*}

\begin{table*}[htbp]
\centering
\resizebox{\textwidth}{!}{%
\begin{tabular}{lcccccc}
\toprule
\multirow{2}{*}{\textbf{Model}} & \multicolumn{3}{c}{\textbf{Recall}} & \multicolumn{3}{c}{\textbf{MRR}} \\
\cmidrule(lr){2-4} \cmidrule(lr){5-7} 
 & \textbf{@10} & \textbf{@100} & \textbf{@1000} & \textbf{@10} & \textbf{@100} & \textbf{@1000} \\
\midrule
\textbf{BGE-M3} & \textbf{0.436} $\pm$ \textbf{0.021} & \textbf{0.566} $\pm$ \textbf{0.017} & 0.723 $\pm$ 0.026 & \textbf{0.402} $\pm$ \textbf{0.026} & \textbf{0.408} $\pm$ \textbf{0.025} & \textbf{0.408} $\pm$ \textbf{0.025} \\
\textbf{E5-large} & 0.434 $\pm$ 0.039 & 0.558 $\pm$ 0.026 & \textbf{0.743} $\pm$ \textbf{0.024} & 0.384 $\pm$ 0.043 & 0.391 $\pm$ 0.043 & 0.391 $\pm$ 0.043 \\
\textbf{E5-base} & 0.419 $\pm$ 0.032 & 0.535 $\pm$ 0.028 & 0.715 $\pm$ 0.021 & 0.377 $\pm$ 0.033 & 0.383 $\pm$ 0.032 & 0.384 $\pm$ 0.032 \\
\textbf{E5-small} & 0.381 $\pm$ 0.031 & 0.503 $\pm$ 0.017 & 0.666 $\pm$ 0.007 & 0.356 $\pm$ 0.029 & 0.362 $\pm$ 0.028 & 0.362 $\pm$ 0.028 \\
\textbf{Granite-278m} & 0.365 $\pm$ 0.032 & 0.469 $\pm$ 0.029 & 0.630 $\pm$ 0.009 & 0.327 $\pm$ 0.018 & 0.333 $\pm$ 0.017 & 0.333 $\pm$ 0.017 \\
\textbf{Granite-107m} & 0.265 $\pm$ 0.029 & 0.353 $\pm$ 0.033 & 0.525 $\pm$ 0.023 & 0.229 $\pm$ 0.009 & 0.233 $\pm$ 0.009 & 0.234 $\pm$ 0.009 \\
\textbf{SPhilBERTa} & 0.358 $\pm$ 0.032 & 0.506 $\pm$ 0.046 & 0.690 $\pm$ 0.028 & 0.303 $\pm$ 0.031 & 0.310 $\pm$ 0.030 & 0.311 $\pm$ 0.030 \\
\bottomrule
\end{tabular}
}
\caption{\textbf{Performance comparison of different base models for the dense retrieval task.} Each held-out query is ranked against the full source corpus of 90,546 segments; Recall@$k$ and MRR@$k$ are computed per query with both reference types pooled (mean~$\pm$~std over 5 folds). All bi-encoders were fine-tuned with the same negative-sampling strategy and training schedule (Online Contrastive Loss, 4 epochs, learning rate $2\times10^{-5}$; Appendix~\ref{app:retrieval_setup}). The larger multilingual encoders (BGE-M3, E5-large) lead on every metric; Table~\ref{tab:summary_results_retrieval} reports the same runs split by reference type. Best value per column in bold.}
\label{tab:ir_model_results}
\end{table*}

\begin{table*}[htbp]
\centering
\resizebox{\textwidth}{!}{%
\begin{tabular}{lcccccc}
\toprule
\multirow{2}{*}{\textbf{Model}} & \multicolumn{3}{c}{\textbf{Recall}} & \multicolumn{3}{c}{\textbf{MRR}} \\
\cmidrule(lr){2-4} \cmidrule(lr){5-7}
 & \textbf{@10} & \textbf{@100} & \textbf{@1000} & \textbf{@10} & \textbf{@100} & \textbf{@1000} \\
\midrule
\textbf{TF-IDF (surf. 1g)} & 0.516 $\pm$ 0.037 & 0.652 $\pm$ 0.024 & 0.804 $\pm$ 0.019 & 0.439 $\pm$ 0.035 & 0.446 $\pm$ 0.034 & 0.446 $\pm$ 0.034 \\
\textbf{TF-IDF (lem. 1g)} & 0.475 $\pm$ 0.050 & 0.623 $\pm$ 0.027 & 0.821 $\pm$ 0.014 & 0.431 $\pm$ 0.048 & 0.438 $\pm$ 0.047 & 0.439 $\pm$ 0.046 \\
\textbf{TF-IDF (lem. 1+2g)} & 0.494 $\pm$ 0.039 & 0.645 $\pm$ 0.040 & \textbf{0.833} $\pm$ \textbf{0.021} & 0.419 $\pm$ 0.037 & 0.426 $\pm$ 0.037 & 0.426 $\pm$ 0.036 \\
\textbf{BM25 (surf. 1g)} & \textbf{0.572} $\pm$ \textbf{0.041} & 0.666 $\pm$ 0.020 & 0.799 $\pm$ 0.020 & \textbf{0.556} $\pm$ \textbf{0.032} & \textbf{0.560} $\pm$ \textbf{0.031} & \textbf{0.561} $\pm$ \textbf{0.031} \\
\textbf{BM25 (lem. 1g)} & 0.545 $\pm$ 0.046 & \textbf{0.680} $\pm$ \textbf{0.011} & 0.824 $\pm$ 0.007 & 0.514 $\pm$ 0.056 & 0.521 $\pm$ 0.053 & 0.522 $\pm$ 0.053 \\
\textbf{Word2Vec (WE, 2g)} & 0.122 $\pm$ 0.012 & 0.264 $\pm$ 0.023 & 0.529 $\pm$ 0.033 & 0.077 $\pm$ 0.005 & 0.083 $\pm$ 0.006 & 0.084 $\pm$ 0.006 \\
\textbf{fastText (WE, LiLa)} & 0.131 $\pm$ 0.017 & 0.280 $\pm$ 0.042 & 0.517 $\pm$ 0.038 & 0.073 $\pm$ 0.015 & 0.079 $\pm$ 0.014 & 0.080 $\pm$ 0.014 \\
\textbf{LatinBERT (WE)} & 0.485 $\pm$ 0.025 & 0.616 $\pm$ 0.022 & 0.765 $\pm$ 0.009 & 0.452 $\pm$ 0.032 & 0.458 $\pm$ 0.032 & 0.459 $\pm$ 0.032 \\
\bottomrule
\end{tabular}
}
\caption{\textbf{Performance comparison of baseline retrieval approaches.} Same protocol as Table~\ref{tab:ir_model_results}: full source corpus, both reference types pooled, mean~$\pm$~std over 5 folds. Lexical approaches (TF-IDF, BM25) use surface tokens (\textit{surf.}) or CLTK lemmas (\textit{lem.}) with unigrams (\textit{1g}) or unigrams+bigrams (\textit{1+2g}) and have no learned parameters. WE denotes word-embedding baselines: bigram Word2Vec vectors following \citet{DBLP:conf/naacl/BurnsBLCD21}, a LiLa fastText checkpoint, and position-aggregated LatinBERT token embeddings following \citet{gong_augmented_2025} (tokenization: Appendix~\ref{app:retrieval_setup}). BM25 over lemma unigrams gives the best Recall@100 and is the first stage of the pipeline; the LatinBERT embeddings are the strongest neural retriever, ahead of every bi-encoder in Table~\ref{tab:ir_model_results}. Best value per column in bold.}
\label{tab:ir_baseline_results}
\end{table*}

\section{Classification Results}
\label{app:classification_results}

We evaluate the multiclass cross-encoder with standard classification metrics and the task-specific global error rates defined in Appendix~\ref{app:metrics}.

\subsection{Sampling and Models}
\label{app:classification_main_results}

Table~\ref{tab:sampling_results} shows that query-conditioned random negatives ($\langle \text{qry}, \text{rnd} \rangle$) reduce global error rates more effectively than fully random pairs ($\langle \text{rnd}, \text{rnd} \rangle$). Table~\ref{tab:classification_sampling_proportion_3class} reports the 3-class model under different class-sampling proportions. Table~\ref{tab:model_results} compares the base architectures, while Figure~\ref{fig:neg_sample_proportion} summarizes the negative-ratio sweep.

Three patterns stand out. First, the composition of the negatives matters more than their number: mixed negatives ($\langle \text{qry}, \text{mix} \rangle$) yield the smallest global error (SMR 0.045) and the fewest false positives (7,084), query-conditioned random negatives come next (SMR 0.063), whereas fully random pairs let the false positives explode to 44,578 (SMR 0.289) because the model never sees a semantically plausible distractor during training; hard negatives alone (SMR 0.111) also fall short of the mixture, so easy negatives remain necessary to calibrate the decision boundary. Second, raising the share of negatives from $1{:}1$ to $1{:}10$ lowers the false-positive rate monotonically (FPR 0.19 to 0.03) at a modest cost in recall (0.67 to 0.61; Table~\ref{tab:classification_sampling_proportion_3class}), which is the trade-off behind the $1{:}10$ default. Third, among the encoders (Table~\ref{tab:model_results}) LatinBERT reaches the highest verbatim F1 (0.78) and macro F1 (0.45), within a fold std of XLM-R Large (0.76, 0.43) and mmBERT Base (0.74, 0.43); mmBERT Small has the best allusion F1 (0.13), BERT-Romanian (0.63) matches PhilBERTa (0.64), and LaBERTa (0.32) and RoBERTa-Latin (0.02) fall behind, and the lexical classifiers combine the highest verbatim recall (GBDT 0.84) with precision below 0.4.

\begin{table*}[htbp]
\centering
\resizebox{\textwidth}{!}{%
\begin{tabular}{lccccccccrrr}
\toprule
\multirow{2}{*}{\textbf{Sampling Method}} & \multicolumn{4}{c}{\textbf{Classification Metrics}} & \multicolumn{3}{c}{\textbf{Global Error Rates}} & \multicolumn{4}{c}{\textbf{Confusion Matrix}} \\
\cmidrule(lr){2-5} \cmidrule(lr){6-8} \cmidrule(lr){9-12}
 & \textbf{Prec.} & \textbf{Rec.} & \textbf{F1} & \textbf{Acc.} & \textbf{FPR} & \textbf{FNR} & \textbf{SMR} & \textbf{TP} & \textbf{FP} & \textbf{FN} & \textbf{TN} \\
\midrule
\textbf{Mixed negatives} $\langle \text{qry}, \text{mix} \rangle$ & \textbf{0.03} & 0.61 & \textbf{0.05} & \textbf{0.96} & \textbf{0.0443} & 0.0007 & \textbf{0.0450} & 179 & \textbf{7084} & 114 & \textbf{152150} \\
\textbf{Random pairs} $\langle \text{rnd}, \text{rnd} \rangle$ & 0.01 & \textbf{0.67} & 0.01 & 0.71 & 0.2888 & \textbf{0.0006} & 0.2894 & \textbf{195} & 44578 & \textbf{99} & 114661 \\
\textbf{Random negatives} $\langle \text{qry}, \text{rnd} \rangle$ & 0.02 & 0.66 & 0.04 & 0.94 & 0.0626 & 0.0006 & 0.0633 & 193 & 10154 & 101 & 149085 \\
\textbf{Hard negatives} $\langle \text{qry}, \text{sim} \rangle$ & 0.01 & 0.61 & 0.02 & 0.89 & 0.1100 & 0.0007 & 0.1107 & 179 & 17554 & 114 & 141682 \\
\bottomrule
\end{tabular}
}
\caption{\textbf{Performance comparison of the classification model across different negative sampling strategies.} Metrics are computed from the thresholded multiclass evaluation. The 3-class confusion matrix is collapsed over the two positive classes (\textit{cit.} and \textit{cf.}); wrong positive-type predictions count as class-level errors. FPR, FNR, and SMR denote global error rates normalized by the total number of pairs $N$. Results are averaged across 5 folds. Best value per column in bold.}
\label{tab:sampling_results}
\end{table*}

\setcounter{topnumber}{4}\setcounter{totalnumber}{5}\setcounter{dbltopnumber}{4}
\renewcommand{\topfraction}{.95}\renewcommand{\textfraction}{.05}
\renewcommand{\dbltopfraction}{.95}\renewcommand{\dblfloatpagefraction}{.5}
\renewcommand{\floatpagefraction}{.8}

\begin{table*}[p]
\centering
\resizebox{\textwidth}{!}{%
\begin{tabular}{lccccccccrrr}
\toprule
\multirow{2}{*}{\textbf{Sampling Proportion}} & \multicolumn{4}{c}{\textbf{Classification Metrics}} & \multicolumn{3}{c}{\textbf{Global Error Rates}} & \multicolumn{4}{c}{\textbf{Confusion Matrix}} \\
\cmidrule(lr){2-5} \cmidrule(lr){6-8} \cmidrule(lr){9-12}
 & \textbf{Prec.} & \textbf{Rec.} & \textbf{F1} & \textbf{Acc.} & \textbf{FPR} & \textbf{FNR} & \textbf{SMR} & \textbf{TP} & \textbf{FP} & \textbf{FN} & \textbf{TN} \\
\midrule
\textbf{$1:1$ neg. \{0: 0.50, 1: 0.29, 2: 0.21\}} & 0.01 & 0.67 & 0.02 & 0.81 & 0.1917 & 0.0006 & 0.1924 & 195 & 30148 & 98 & 129099 \\
\textbf{$1:2$ neg. \{0: 0.67, 1: 0.19, 2: 0.14\}} & 0.01 & 0.69 & 0.02 & 0.83 & 0.1704 & 0.0006 & 0.1709 & 202 & 27353 & 91 & 131896 \\
\textbf{$1:3$ neg. \{0: 0.75, 1: 0.14, 2: 0.11\}} & 0.03 & 0.64 & 0.07 & 0.96 & 0.0401 & 0.0007 & 0.0407 & 188 & 6340 & 106 & 152910 \\
\textbf{$1:4$ neg. \{0: 0.80, 1: 0.12, 2: 0.08\}} & 0.03 & \textbf{0.70} & 0.05 & 0.94 & 0.0640 & \textbf{0.0006} & 0.0646 & \textbf{205} & 10643 & \textbf{88} & 148601 \\
\textbf{$1:5$ neg. \{0: 0.83, 1: 0.10, 2: 0.07\}} & 0.02 & 0.68 & 0.03 & 0.91 & 0.0924 & 0.0006 & 0.0930 & 200 & 14984 & 93 & 144253 \\
\textbf{$1:6$ neg. \{0: 0.86, 1: 0.08, 2: 0.06\}} & 0.03 & 0.65 & 0.06 & 0.96 & 0.0396 & 0.0006 & 0.0402 & 191 & 6177 & 102 & 153063 \\
\textbf{$1:7$ neg. \{0: 0.88, 1: 0.07, 2: 0.05\}} & 0.02 & 0.66 & 0.04 & 0.94 & 0.0590 & 0.0006 & 0.0596 & 194 & 9520 & 100 & 149714 \\
\textbf{$1:8$ neg. \{0: 0.89, 1: 0.06, 2: 0.05\}} & 0.04 & 0.66 & 0.07 & 0.96 & 0.0373 & 0.0006 & 0.0380 & 193 & 5894 & 101 & 153345 \\
\textbf{$1:9$ neg. \{0: 0.90, 1: 0.06, 2: 0.04\}} & 0.03 & 0.63 & 0.06 & 0.95 & 0.0480 & 0.0007 & 0.0487 & 184 & 7491 & 109 & 151743 \\
\textbf{$1:10$ neg. \{0: 0.91, 1: 0.05, 2: 0.04\}} & \textbf{0.04} & 0.61 & \textbf{0.08} & \textbf{0.97} & \textbf{0.0335} & 0.0007 & \textbf{0.0342} & 179 & \textbf{5245} & 115 & \textbf{153993} \\
\bottomrule
\end{tabular}
}
\caption{\textbf{Performance comparison of the 3-class classification model across different class-sampling proportions.} Metrics are computed from the thresholded multiclass evaluation. The 3-class confusion matrix is collapsed over the two positive classes (\textit{cit.} and \textit{cf.}); wrong positive-type predictions count as class-level errors. FPR, FNR, and SMR denote global error rates normalized by the total number of pairs $N$. Results are averaged across folds. Best value per column in bold.}
\label{tab:classification_sampling_proportion_3class}
\end{table*}

\begin{table*}[p]
\centering
\resizebox{\textwidth}{!}{%
\begin{tabular}{l@{\hspace{15pt}}ccc@{\hspace{15pt}}ccc@{\hspace{15pt}}c}
\toprule
\multirow{2}{*}{\textbf{Model}} & \multicolumn{3}{c}{\textbf{Verbatim (cit.)}} & \multicolumn{3}{c}{\textbf{Allusion (cf.)}} & \multirow{2}{*}{\textbf{Avg. F1}} \\
\cmidrule(lr){2-4}\cmidrule(lr){5-7}
 & \textbf{Prec.} & \textbf{Rec.} & \textbf{F1} & \textbf{Prec.} & \textbf{Rec.} & \textbf{F1} & \\
\midrule
\textbf{LatinBERT} & $\smash{\underline{\mathbf{0.77 \pm 0.04}}}$ & $0.83 \pm 0.03$ & $\smash{\underline{\mathbf{0.78 \pm 0.03}}}$ & $0.08 \pm 0.03$ & $0.42 \pm 0.08$ & $0.11 \pm 0.03$ & $\smash{\underline{\mathbf{0.45 \pm 0.03}}}$ \\
\textbf{mmBERT Base} & $0.75 \pm 0.02$ & $0.75 \pm 0.02$ & $0.74 \pm 0.02$ & $0.09 \pm 0.03$ & $0.45 \pm 0.08$ & $0.13 \pm 0.04$ & $0.43 \pm 0.02$ \\
\textbf{mmBERT Small} & $0.72 \pm 0.07$ & $0.77 \pm 0.04$ & $0.72 \pm 0.06$ & $\smash{\underline{\mathbf{0.09 \pm 0.02}}}$ & $0.37 \pm 0.05$ & $\smash{\underline{\mathbf{0.13 \pm 0.03}}}$ & $0.42 \pm 0.04$ \\
\textbf{XLM-R Large} & $0.75 \pm 0.06$ & $0.81 \pm 0.04$ & $0.76 \pm 0.04$ & $0.07 \pm 0.01$ & $0.53 \pm 0.09$ & $0.11 \pm 0.01$ & $0.43 \pm 0.02$ \\
\textbf{XLM-R Base} & $0.69 \pm 0.10$ & $0.75 \pm 0.04$ & $0.69 \pm 0.08$ & $0.07 \pm 0.02$ & $\smash{\underline{\mathbf{0.56 \pm 0.08}}}$ & $0.11 \pm 0.03$ & $0.40 \pm 0.05$ \\
\textbf{mBERT Large} & $0.71 \pm 0.03$ & $0.78 \pm 0.03$ & $0.72 \pm 0.02$ & $0.08 \pm 0.01$ & $0.36 \pm 0.08$ & $0.12 \pm 0.02$ & $0.42 \pm 0.01$ \\
\textbf{mBERT Base} & $0.69 \pm 0.07$ & $0.75 \pm 0.03$ & $0.69 \pm 0.05$ & $0.08 \pm 0.02$ & $0.42 \pm 0.02$ & $0.12 \pm 0.02$ & $0.41 \pm 0.04$ \\
\textbf{BERT-Romanian} & $0.62 \pm 0.03$ & $0.71 \pm 0.03$ & $0.63 \pm 0.02$ & $0.07 \pm 0.03$ & $0.42 \pm 0.06$ & $0.11 \pm 0.04$ & $0.37 \pm 0.02$ \\
\textbf{PhilBERTa} & $0.62 \pm 0.09$ & $0.73 \pm 0.04$ & $0.64 \pm 0.07$ & $0.03 \pm 0.01$ & $0.45 \pm 0.10$ & $0.05 \pm 0.02$ & $0.34 \pm 0.04$ \\
\textbf{LaBERTa} & $0.25 \pm 0.17$ & $0.66 \pm 0.05$ & $0.32 \pm 0.15$ & $0.02 \pm 0.01$ & $0.24 \pm 0.06$ & $0.03 \pm 0.01$ & $0.17 \pm 0.08$ \\
\textbf{RoBERTa-Latin} & $0.01 \pm 0.00$ & $0.26 \pm 0.03$ & $0.02 \pm 0.00$ & $0.01 \pm 0.00$ & $0.12 \pm 0.04$ & $0.02 \pm 0.00$ & $0.02 \pm 0.00$ \\
\midrule
\textbf{LogReg (lexical)} & $0.37 \pm 0.06$ & $0.83 \pm 0.03$ & $0.46 \pm 0.06$ & $0.02 \pm 0.01$ & $0.37 \pm 0.07$ & $0.04 \pm 0.01$ & $0.25 \pm 0.03$ \\
\textbf{GBDT (lexical)} & $0.36 \pm 0.04$ & $\smash{\underline{\mathbf{0.84 \pm 0.02}}}$ & $0.45 \pm 0.04$ & $0.02 \pm 0.01$ & $0.22 \pm 0.07$ & $0.04 \pm 0.01$ & $0.24 \pm 0.02$ \\
\bottomrule
\end{tabular}
}
\caption{\textbf{Performance comparison of different pre-trained base models used in the classification stage.} Scores are given for \emph{Verbatim} (\textit{cit.}) and \emph{Allusion} (\textit{cf.}); the last column is the macro-averaged F1. Per-class decision thresholds are tuned on the training split of each fold. Values are mean~$\pm$~std over 5 cross-validation folds. The best value per column is bold and underlined.}
\label{tab:model_results}
\end{table*}

\begin{table*}[p]
\centering
\resizebox{\textwidth}{!}{%
\begin{tabular}{rrrrrrrrrrrrrrrr}
\toprule
 & \multicolumn{2}{c}{\textbf{\# Predictions}} & \multicolumn{4}{c}{\textbf{Retrieval Only}} & \multicolumn{4}{c}{\textbf{Retrieve+Rerank}} & \multicolumn{2}{c}{\textbf{FPR $\times 10^4$}} & \multicolumn{2}{c}{\textbf{SMR $\times 10^4$}} \\
\cmidrule(lr){2-3} \cmidrule(lr){4-7} \cmidrule(lr){8-11} \cmidrule(lr){12-13} \cmidrule(lr){14-15}
\textbf{$k$} & \textbf{Ret.} & \textbf{Rer.} & \textbf{TP} & \textbf{FP} & \textbf{FN} & \textbf{F1} & \textbf{TP} & \textbf{FP} & \textbf{FN} & \textbf{F1} & \textbf{Ret.} & \textbf{Rer.} & \textbf{Ret.} & \textbf{Rer.} \\
\midrule
5 & 1{,}412 & 321 & 164 & 1{,}248 & 130 & 0.19 & 127 & 193 & 167 & 0.42 & 39.42 & 6.07 & 43.52 & 11.28 \\
10 & 2{,}824 & 503 & 189 & 2{,}635 & 105 & 0.12 & 142 & 361 & 152 & 0.36 & 83.27 & 11.30 & 86.60 & 16.05 \\
20 & 5{,}648 & 777 & 215 & 5{,}433 & 79 & 0.07 & 157 & 620 & 137 & 0.30 & 171.66 & 19.37 & 174.15 & 23.66 \\
50 & 14{,}120 & 1{,}355 & 250 & 13{,}870 & 44 & 0.03 & 176 & 1{,}179 & 118 & 0.22 & 438.26 & 36.76 & 439.65 & 40.43 \\
100 & 28{,}240 & 1{,}901 & 269 & 27{,}971 & 25 & 0.02 & 185 & 1{,}715 & 109 & 0.17 & 883.81 & 53.41 & 884.59 & 56.78 \\
500 & 141{,}200 & 3{,}253 & 294 & 140{,}906 & 0 & 0.00 & 197 & 3{,}056 & 97 & 0.12 & 4452.32 & 94.03 & 4452.34 & 97.00 \\
1000 & 159{,}489 & 3{,}347 & 294 & 159{,}195 & 0 & 0.00 & 197 & 3{,}150 & 97 & 0.12 & 5030.20 & 96.80 & 5030.20 & 99.76 \\
\bottomrule
\end{tabular}
}
\caption{\textbf{Comparison between retrieval-only and retrieve-and-rerank (best retrieve-and-rerank configuration: BM25 + XLM-R Large).} $k$ denotes the number of most similar candidates retrieved by BM25 that are reranked using the classification model. \textbf{Ret.} = total top-$k$ candidates retrieved (= $k \times n_\text{queries}$); \textbf{Rer.} = positive predictions output by the reranker. Results are pooled across verbatim and allusion reference types and averaged across 5 folds. The reranking stage substantially reduces error rates (FPR, SMR), scaled by $\times 10^4$ for readability.}
\label{tab:pipeline_comparison}
\end{table*}

\setcounter{topnumber}{2}\setcounter{totalnumber}{3}\setcounter{dbltopnumber}{2}
\renewcommand{\topfraction}{.7}\renewcommand{\textfraction}{.2}
\renewcommand{\dbltopfraction}{.7}\renewcommand{\dblfloatpagefraction}{.5}
\renewcommand{\floatpagefraction}{.5}

\section{Retrieve-and-Rerank Results}
\label{app:combined_results}

Table~\ref{tab:pipeline_comparison} compares the retrieval-only baseline with the retrieve-and-rerank pipeline at varying retrieval depths ($k$), relating performance gains to the number of candidate pairs sent to the classifier. Figure~\ref{fig:pipeline_tradeoff} visualizes the same precision--efficiency trade-off.

Two observations follow from Table~\ref{tab:pipeline_comparison}. At $k=5$ the reranker keeps 321 of the 1,412 retrieved candidates, retaining 127 of the 164 true references in the pool while cutting the false positives from 1,248 to 193 (FPR $\times 10^4$ from 39.4 to 6.1). As $k$ grows, the retrieval-only baseline reaches all 294 references only at $k \geq 500$, at the price of more than 140,000 false positives, whereas the pipeline's false positives grow slowly (3,056 at $k=500$) and its recall saturates at 197 of 294 references (67\%): the remaining 97 are retrieved but rejected by the classifier, i.e., they are the classifier's own misses analyzed in Appendix~\ref{app:error_analysis_details}. Beyond $k \approx 100$ the additional candidates add almost no true positives for the pipeline (185 to 197) but nearly double the number of flagged pairs (1,901 to 3,347), which is why the main text reports $k=100$ as the practical operating point. Figure~\ref{fig:pipeline_tradeoff} shows the same trade-off for the two alternative configurations: replacing BM25 by BGE-M3 as the first stage lowers recall at every depth, in line with Table~\ref{tab:summary_results_retrieval}, and replacing XLM-R Large by the LogReg reranker keeps recall but forfeits most of the precision gain.

%% file: sections/appendix/11b_error_analysis.tex
\newcommand{\querymarker}{\textcolor{userprompttitletext}{\raisebox{-0.15ex}{\tikz[scale=0.11, line width=0.7pt]{\draw (0,0) circle (1); \draw (0.72,-0.72) -- (1.55,-1.55);}}}}
\newcommand{\correctmarker}{\textcolor{green!45!black}{\ding{51}}}
\newcommand{\wrongmarker}{\textcolor{red!75!black}{\ding{55}}}
\section{Error Analysis}
\makeatletter\global\@colroom\@colht\pagegoal\@colht\makeatother

\label{app:error_analysis_details}

\begin{figure*}[p]
\centering
\includegraphics[width=0.88\linewidth]{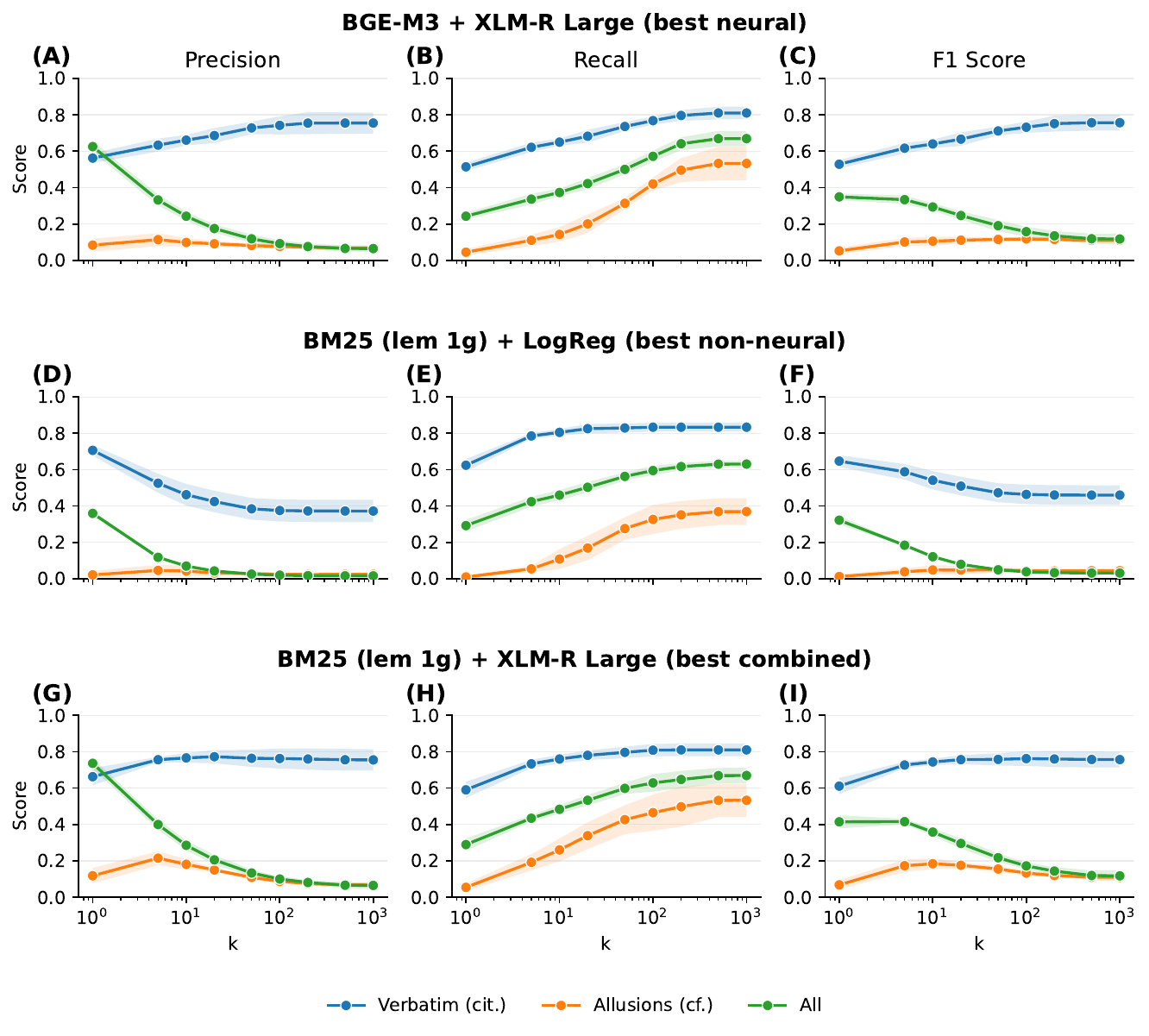}
\captionof{figure}{\textbf{Precision, Recall, and F1 Score of three retrieve-and-rerank pipeline configurations} as a function of the number of retrieved candidates~$k$. Each row corresponds to one configuration: (\textit{top}) BGE-M3 retrieval + XLM-R Large reranker, (\textit{middle}) BM25 retrieval + LogReg reranker, (\textit{bottom}) BM25 retrieval + XLM-R Large reranker. Within each panel, results are reported separately for verbatim references (\textit{cit.}), allusions (\textit{cf.}), and all references pooled (\textit{all}). Shaded bands indicate $\pm1$ standard deviation across 5 cross-validation folds.}
\label{fig:pipeline_tradeoff}
\vspace{0.8cm}
\centering
\resizebox{\textwidth}{!}{%
\begin{tabular}{llcccccccccccccccc}
\toprule
\multirow{2}{*}{\textbf{Query}} & \multirow{2}{*}{\textbf{Source}} & \multicolumn{8}{c}{\textbf{Verbatim (cit.)}} & \multicolumn{8}{c}{\textbf{Allusion (cf.)}} \\
\cmidrule(lr){3-10} \cmidrule(lr){11-18}
 & & \textbf{\#} & \textbf{Ov.} & \textbf{P} & \textbf{R} & \textbf{F1} & \textbf{FPR} & \textbf{FNR} & \textbf{SMR} & \textbf{\#} & \textbf{Ov.} & \textbf{P} & \textbf{R} & \textbf{F1} & \textbf{FPR} & \textbf{FNR} & \textbf{SMR} \\
\midrule
Jerome & Cicero & 91 & \ov{5.22} & \f{0.651} & \f{0.673} & \f{0.656} & \err{0.003} & \err{0.000} & \err{0.004} & 21 & \ov{3.14} & \f{0.050} & \f{0.033} & \f{0.040} & \err{0.001} & \err{0.000} & \err{0.001} \\
Jerome & Horace & 41 & \ov{6.80} & \f{0.980} & \f{0.980} & \f{0.980} & \err{0.003} & \err{0.000} & \err{0.003} & 0 & -- & -- & -- & -- & -- & -- & -- \\
Jerome & Lucan & 2 & \ov{5.00} & \f{1.000} & \f{1.000} & \f{1.000} & \err{0.003} & \err{0.000} & \err{0.003} & 0 & -- & -- & -- & -- & -- & -- & -- \\
Jerome & Ovid & 2 & \ov{6.00} & \f{1.000} & \f{1.000} & \f{1.000} & \err{0.003} & \err{0.000} & \err{0.003} & 0 & -- & -- & -- & -- & -- & -- & -- \\
Jerome & Virgil & 211 & \ov{6.95} & \f{0.878} & \f{0.944} & \f{0.894} & \err{0.004} & \err{0.000} & \err{0.004} & 38 & \ov{3.20} & \f{0.077} & \f{0.474} & \f{0.117} & \err{0.017} & \err{0.000} & \err{0.018} \\
Lactantius & Horace & 5 & \ov{12.40} & \f{1.000} & \f{1.000} & \f{1.000} & \err{0.000} & \err{0.000} & \err{0.000} & 0 & -- & -- & -- & -- & -- & -- & -- \\
Lactantius & Lucretius & 24 & \ov{13.21} & \f{0.988} & \f{1.000} & \f{0.992} & \err{0.001} & \err{0.000} & \err{0.001} & 0 & -- & -- & -- & -- & -- & -- & -- \\
Lactantius & Ovid & 23 & \ov{11.13} & \f{0.980} & \f{1.000} & \f{0.987} & \err{0.001} & \err{0.000} & \err{0.001} & 0 & -- & -- & -- & -- & -- & -- & -- \\
Lactantius & Propertius & 3 & \ov{8.00} & \f{1.000} & \f{1.000} & \f{1.000} & \err{0.000} & \err{0.000} & \err{0.000} & 0 & -- & -- & -- & -- & -- & -- & -- \\
Lactantius & Virgil & 78 & \ov{8.85} & \f{0.894} & \f{0.983} & \f{0.917} & \err{0.001} & \err{0.000} & \err{0.001} & 0 & -- & -- & -- & -- & -- & -- & -- \\
Valerius & Lucan & 65 & \ov{1.91} & \f{0.529} & \f{0.522} & \f{0.518} & \err{0.002} & \err{0.003} & \err{0.005} & 85 & \ov{1.33} & \f{0.236} & \f{0.641} & \f{0.324} & \err{0.078} & \err{0.003} & \err{0.081} \\
Valerius & Ovid & 55 & \ov{1.86} & \f{0.540} & \f{0.511} & \f{0.521} & \err{0.001} & \err{0.001} & \err{0.002} & 94 & \ov{1.16} & \f{0.214} & \f{0.609} & \f{0.294} & \err{0.047} & \err{0.002} & \err{0.049} \\
Valerius & Statius & 42 & \ov{2.26} & \f{0.490} & \f{0.500} & \f{0.492} & \err{0.004} & \err{0.003} & \err{0.007} & 78 & \ov{1.70} & \f{0.163} & \f{0.570} & \f{0.234} & \err{0.112} & \err{0.004} & \err{0.116} \\
Valerius & Virgil & 207 & \ov{2.13} & \f{0.552} & \f{0.545} & \f{0.527} & \err{0.002} & \err{0.002} & \err{0.004} & 305 & \ov{1.57} & \f{0.089} & \f{0.602} & \f{0.144} & \err{0.086} & \err{0.003} & \err{0.089} \\
\midrule
Total &  & 849 & \ov{5.11} & \f{0.755} & \f{0.810} & \f{0.757} & \err{0.003} & \err{0.000} & \err{0.003} & 621 & \ov{1.65} & \f{0.067} & \f{0.533} & \f{0.110} & \err{0.017} & \err{0.000} & \err{0.017} \\
\bottomrule
\end{tabular}
}
\captionof{table}{\textbf{Per-pair, per-class breakdown for XLM-R Large on the full pairwise evaluation set (mean across 5 folds).} Every rate metric (P, R, F1, FPR, FNR, SMR) is computed per fold using the \texttt{locisimiles} per-query macro definition (one-vs-rest at $p\geq 0.5$); support (\#) is the sum across folds. Performance metrics (P, R, F1) are shaded teal (darker = larger), error rates (FPR, FNR, SMR) and overlap are shaded so that darker indicates larger values.}
\label{tab:xlmr_per_pair_class_fullgrid_mean}
\end{figure*}

Table~\ref{tab:xlmr_per_pair_class_fullgrid_mean} reports the full per-author-pair, per-class breakdown for XLM-R Large on the pairwise evaluation grid. Unlike the compact table in the main error-analysis section, this appendix table includes precision, recall, F1, and the \texttt{locisimiles} error-rate metrics (FPR, FNR, SMR) for each author pair and each positive class.

The table should be read as a descriptive diagnostic, not as a pooled confusion matrix over the entire dataset. Each rate metric is computed within each fold using the \texttt{locisimiles} per-query macro definition in a one-vs-rest setup at $p\geq 0.5$, and the reported values are then averaged over folds. Support (\#), by contrast, is summed across folds. Rows with very small support can therefore show extreme scores; the goal is to expose variation by author pair and reference type rather than to provide a second aggregate leaderboard.

The author-pair breakdown sharpens the pattern from the main text: verbatim performance tracks lexical overlap closely, while allusion performance is weaker and varies more by author pair. Jerome's prose allusions remain especially difficult. The Valerius allusion rows, despite much lower overlap, benefit from larger support and a more homogeneous poetic source domain.

\subsection{Qualitative Error Examples}
\label{subsec:qualitative_analysis}

To complement the quantitative breakdown, we inspect two representative cases in detail: a false positive labeled \textit{cit.}\ and a false negative missed entirely by the model.

\paragraph{False positive.}
Consider the false positive in Figure~\ref{fig:error_examples} (left). The lexical overlap is \textit{omnia timeo} (``I fear everything''), and the verb \textit{timere} (``to fear'') recurs in the source as \textit{timenda sint} (``are to be feared''), strengthening the apparent signal. Yet \textit{omnia timeo} is a very general formulation, carrying little distinctive content on its own. The failure is subtler: Jerome writes \textit{ut ait gentilis poeta} (``as the pagan poet says''), signaling a genuine citation, yet it points to Virgil's more specific \textit{omnia tuta timens} (``fearing all things, even safe ones''). The crucial word \textit{tuta} (``safe things''), absent from the matched source, is exactly what makes the Virgilian phrase distinctive. The model therefore identifies lexically significant overlap effectively, but it illustrates an edge case requiring expert analysis: the overlap points not to the proposed source but to a different one.

\paragraph{False negative.}
Inspecting a case missed entirely by the model (Figure~\ref{fig:error_examples}, right) shows that divergent sentence context can be a problem. The source sentence is wholly concerned with the Socratic saying \textit{nosce te} (``know yourself''). In the query, this reference appears only in the opening imperative \textit{te ipsum intellege!} (``understand yourself!''); the larger part of the query then shifts to a discussion of stylistic simplicity, contrasting \textit{verbosa rusticitas} (``wordy rusticity'') with \textit{sancta simplicitas} (``holy simplicity''). The intertextual signal is therefore locally present but embedded in a broader context that points in a different thematic direction, which may explain why the model fails to retrieve the relevant source.

\begin{figure*}[tp]
\renewcommand{\dbltopfraction}{0.95}\renewcommand{\textfraction}{0.05}
	\centering
	\begin{minipage}[t]{0.48\textwidth}
	\vspace{0pt}
	\begin{tcolorbox}[
		enhanced,
		colback=userpromptbg,
		colframe=userpromptborder,
		colbacktitle=userprompttitle,
		coltitle=userprompttitletext,
		arc=2mm,
		boxrule=1pt,
		left=1.25mm,
		right=1.25mm,
		top=0.75mm,
		bottom=0.75mm,
		fonttitle=\bfseries\small,
		title=False Positive: Generic Overlap,
		toptitle=0.3mm,
		bottomtitle=0.3mm,
		drop shadow]
		\footnotesize
		\querymarker\ \textbf{Query} (\textit{Jerome, Epist. 7.4.1})\\[1pt]
		\textbf{LA:} \textit{huic ego, ut ait gentilis poeta, \textbf{omnia} etiam tuta \textbf{timeo}.}\par
		\textbf{EN:} ``For this reason, as the pagan poet says, \textbf{I fear everything}, even what is safe.''

		\vspace{0.25cm}
		\wrongmarker\ \textbf{Predicted source} (\textit{Cicero, Att. 3.8.2})\\[1pt]
		\textbf{LA:} \textit{reliqua quam mihi timenda sint video nec quid scribam habeo et \textbf{omnia timeo}, nec tam miserum est quicquam quod non in nostram fortunam cadere videatur.}\par
		\vspace{0.10cm}
		\textbf{EN:} ``I see how much I should fear the remaining things, and I do not know what to write; \textbf{I fear everything}, and there is nothing so wretched that it does not seem capable of befalling our lot.''

		\vspace{0.25cm}
		\correctmarker\ \textbf{Actual source} (\textit{Virgil, Aen. 4.298})\\[1pt]
		\textbf{LA:} \textit{praesensit, motusque excepit prima futuros, \textbf{omnia tuta timens}.}\par
		\textbf{EN:} ``She foresaw it, and was the first to detect the coming disturbances, \textbf{fearing everything, even what was safe}.''
		\vspace{0.25cm}
	\end{tcolorbox}
	\end{minipage}\hfill
	\begin{minipage}[t]{0.48\textwidth}
	\vspace{0pt}
	\begin{tcolorbox}[
		enhanced,
		colback=userpromptbg,
		colframe=userpromptborder,
		colbacktitle=userprompttitle,
		coltitle=userprompttitletext,
		arc=2mm,
		boxrule=1pt,
		left=1.25mm,
		right=1.25mm,
		top=0.75mm,
		bottom=0.75mm,
		fonttitle=\bfseries\small,
		title=False Negative: Local Allusion in Broader Context,
		toptitle=0.3mm,
		bottomtitle=0.3mm,
		drop shadow]
		\footnotesize
		\querymarker\ \textbf{Query} (\textit{Jerome, Epist. 57.12.4})\\[1pt]
		\textbf{LA:} \textit{\textbf{te} ipsum \textbf{intellege}! uenerationi mihi semper fuit non uerbosa rusticitas, sed sancta simplicitas: qui in sermone imitari se dicit apostolos, prius imitetur in uita.}\par
		\textbf{EN:} ``\textbf{Understand yourself}! What has always been worthy of reverence to me is not wordy rusticity, but holy simplicity: whoever claims to imitate the apostles in speech should first imitate them in life.''

		\vspace{0.08cm}
		\correctmarker\ \textbf{Source} (\textit{Cicero, Tusc. 1.52})\\[1pt]
		\textbf{LA:} \textit{cum igitur \textbf{nosce te} dicit, hoc dicit: \textbf{nosce animum tuum}.}\par
		\textbf{EN:} ``When, therefore, he says `\textbf{know yourself},' he means this: \textbf{know your own soul}.''
	\end{tcolorbox}
	\caption{\textbf{Qualitative error examples.} Left: false positive labeled \textit{cit.}, where generic \textit{omnia timeo} overlap masks the actual Virgilian source. Right: false negative, where the brief local Socratic allusion \textit{nosce te} is obscured by divergent sentence context.}
	\label{fig:error_examples}
	\end{minipage}

	\vspace{0.6cm}

	\begin{examplebox}[Examples of Intertextual References]
		\small
		\setlength{\parskip}{0.3em}

		\begin{minipage}[t]{0.31\textwidth}
			\centering
			\textbf{ Paraphrase (Minor)}\\
			\par\vspace{0.2cm}
			\raggedright

			\textbf{\textsc{Source}} \\
			\textit{Cicero, Catil. 1.1}
			\begin{quote}
				``Quo usque tandem \textbf{abutere}, Catilina, \textbf{patientia nostra}?'' \\
				\textcolor{gray}{\small (How long, Catiline, will you \textbf{abuse our patience}?)}
			\end{quote}

			\textbf{\textsc{Reuse}} \\
			\textit{Jerome, Epist. 98.22.4}
			\begin{quote}
				``... et \textbf{patientia nostra} quasi quodam temeritatis fomite \textbf{abutentes} ...'' \\
				\textcolor{gray}{\small (... and \textbf{abusing our patience} like some kindling of rashness ...)}
			\end{quote}

			\vfill

			\hrulefill \\
			\vspace{0.1cm}
			\textit{\textbf{Comment:} Jerome integrates Cicero's famous invective syntactically by adapting the verb form (\textit{abutere} $\rightarrow$ \textit{abutentes}).}
		\end{minipage}%
		\hfill\vrule\hfill
		\begin{minipage}[t]{0.31\textwidth}
			\centering
			\textbf{Paraphrase (Major)}\\
			\par\vspace{0.2cm}
			\raggedright

			\textbf{\textsc{Source}} \\
			\textit{Virgil, Georg. 4.82}
			\begin{quote}
				``... \textbf{ingentes animos angusto in pectore} versant ...'' \\
				\textcolor{gray}{\small (... they wield \textbf{mighty souls in a tiny breast} ...)}
			\end{quote}

			\vspace{2em}

			\textbf{\textsc{Reuse}} \\
			\textit{Jerome, Epist. 107.13.4}
			\begin{quote}
				``... et \textbf{in paruis corpusculis ingentes animos} intueri!'' \\
				\textcolor{gray}{\small (... and to see \textbf{mighty souls in small bodies}!)}
			\end{quote}

			\vfill

			\hrulefill \\
			\vspace{0.1cm}
			\textit{\textbf{Comment:} Jerome retains the semantic core but rephrases \textit{angusto in pectore} to \textit{in parvis corpusculis}.}
		\end{minipage}%
		\hfill\vrule\hfill
		\begin{minipage}[t]{0.31\textwidth}
			\centering
			\textbf{Allusion}\\
			\par\vspace{0.2cm}
			\raggedright

			\textbf{\textsc{Source}} \\
			\textit{Cicero, Orat. 33.11}
			\begin{quote}
				``... sed \textbf{nihil difficile amanti} puto.'' \\
				\textcolor{gray}{\small (... but I think \textbf{nothing is difficult for a lover}.)}
			\end{quote}

			\vspace{2em}

			\textbf{\textsc{Reuse}} \\
			\textit{Jerome, Epist. 22.40.1}
			\begin{quote}
				``\textbf{Nihil amantibus durum} est, nullus difficilis cupienti labor.'' \\
				\textcolor{gray}{\small (\textbf{Nothing is hard for lovers}, no labor difficult for the desirous.)}
			\end{quote}

			\vfill

			\hrulefill \\
			\vspace{0.1cm}
			\textit{\textbf{Comment:} Jerome evokes the motif using synonymous but distinct vocabulary (\textit{difficile amanti} vs. \textit{amantibus durum}).}
		\end{minipage}

	\end{examplebox}
	\caption{\textbf{Example references.} Three examples of text reuse by Jerome included in the ground-truth dataset.}
	\label{fig:dataset_examples}
\end{figure*}

%% file: sections/appendix/12_framework.tex
\section{Python Package}
\label{app:software_toolkit}

We release the framework described in this paper as an open-source Python package. The \texttt{locisimiles} library\footnote{\href{https://github.com/julianschelb/locisimiles}{https://github.com/julianschelb/locisimiles}} implements the retrieve-and-rerank pipeline shown in Figure~\ref{fig:two_stage_pipeline} and computes the task-specific error metrics (SMR, FPR, FNR) defined in Appendix~\ref{app:metrics}. It can be installed from PyPI\footnote{\href{https://pypi.org/project/locisimiles/}{https://pypi.org/project/locisimiles/}} with \texttt{pip install locisimiles}, and the API reference is available online.\footnote{\href{https://julianschelb.github.io/locisimiles/}{https://julianschelb.github.io/locisimiles/}}

Because the package compares all segments of the query and source documents, evaluation must account for the correct rejection of pairs labeled \textit{no match} as well as the recovery of references. In addition to per-class precision (P), recall (R), and F1, \texttt{locisimiles} therefore reports three error rates normalized by the total number of segment pairs ($N$): SMR as a global error rate, FPR for spurious matches, and FNR for missed references. Standard accuracy is omitted because the extreme class imbalance makes it uninformative: a trivial all-negative predictor already achieves near-perfect scores. Formal definitions are given in Appendix~\ref{app:metrics}.

\subsection{Python API}
The core API lets researchers load custom query and source documents in CSV format and run the detection pipeline with our fine-tuned models from the Hugging Face Hub.\footnote{\href{https://hf.co/collections/julian-schelb/models-for-latin-intertextuality-search}{https://hf.co/collections/julian-schelb/models-for-latin-intertextuality-search}}

\begin{small}
\noindent\begin{minipage}{\linewidth}
\begin{verbatim}
# 1. Load query and source documents
query_doc = Document("query.csv")
source_doc = Document("source.csv")

# 2. Initialize the pipeline
pipeline = ClassificationPipeline(
    classification_name="...",
)

# 3. Run the pipeline
results = pipeline.run(
    query=query_doc, 
    source=source_doc,
)

# 4. Display results
pretty_print(results)
\end{verbatim}
\end{minipage}
\end{small}

\subsection{Graphical User Interface}

To lower the barrier to entry, the package includes an optional Gradio-based GUI. It can be installed via the optional dependency group (\texttt{pip install .[gui]}) and launched from the command line with \texttt{locisimiles-gui}.

\begin{figure*}[t]
    \centering
    \captionsetup[subfigure]{labelfont=bf, textfont=bf}
    \renewcommand{\thesubfigure}{\Alph{subfigure}}

    \begin{subfigure}[b]{0.48\textwidth}
        \centering
        \caption{Data Upload}
        \label{fig:gui_upload}
        \vspace{5pt}
        \includegraphics[width=\linewidth]{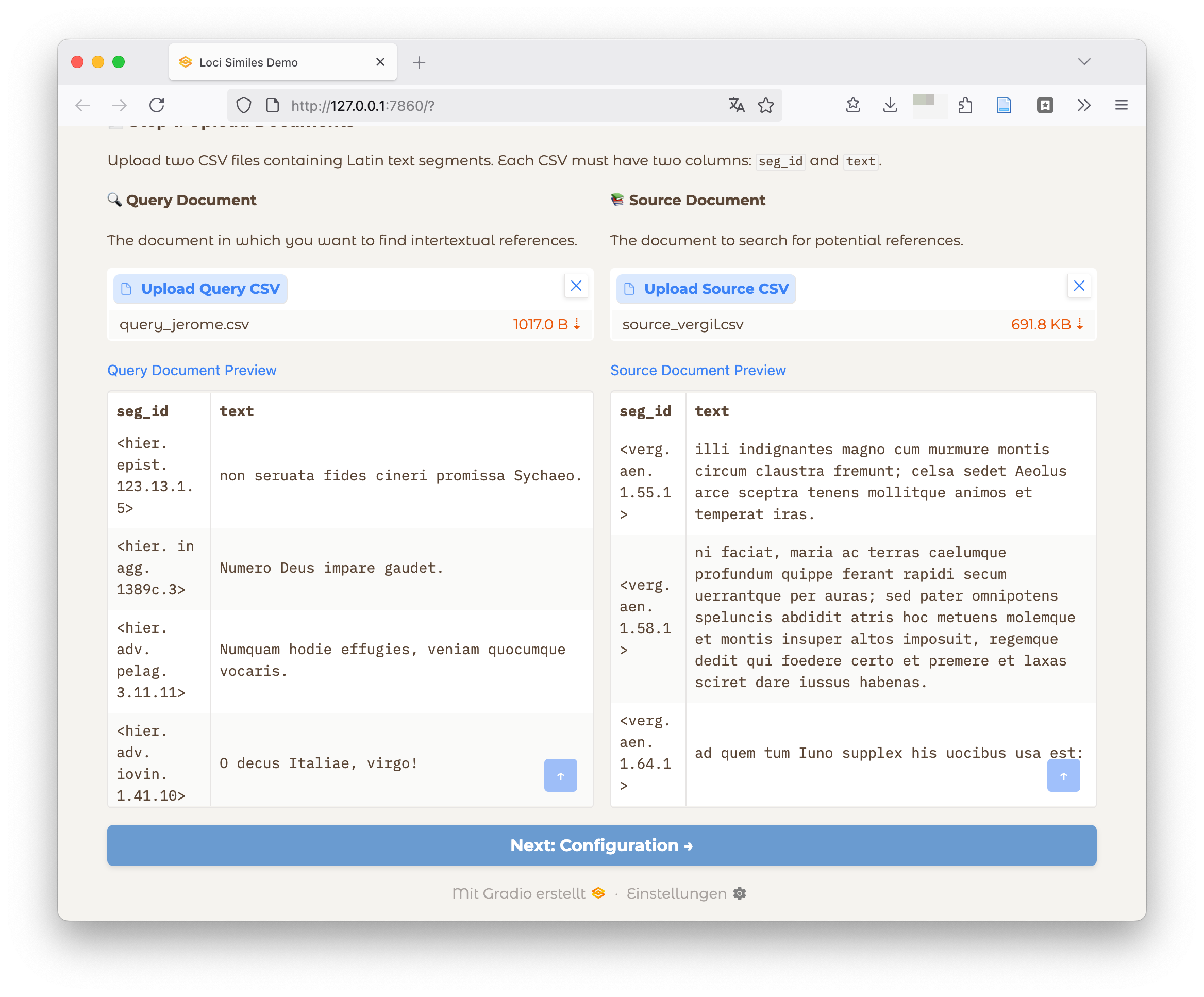}
    \end{subfigure}
    \hfill
    \begin{subfigure}[b]{0.48\textwidth}
        \centering
        \caption{Configuration}
        \label{fig:gui_config}
        \vspace{5pt}
        \includegraphics[width=\linewidth]{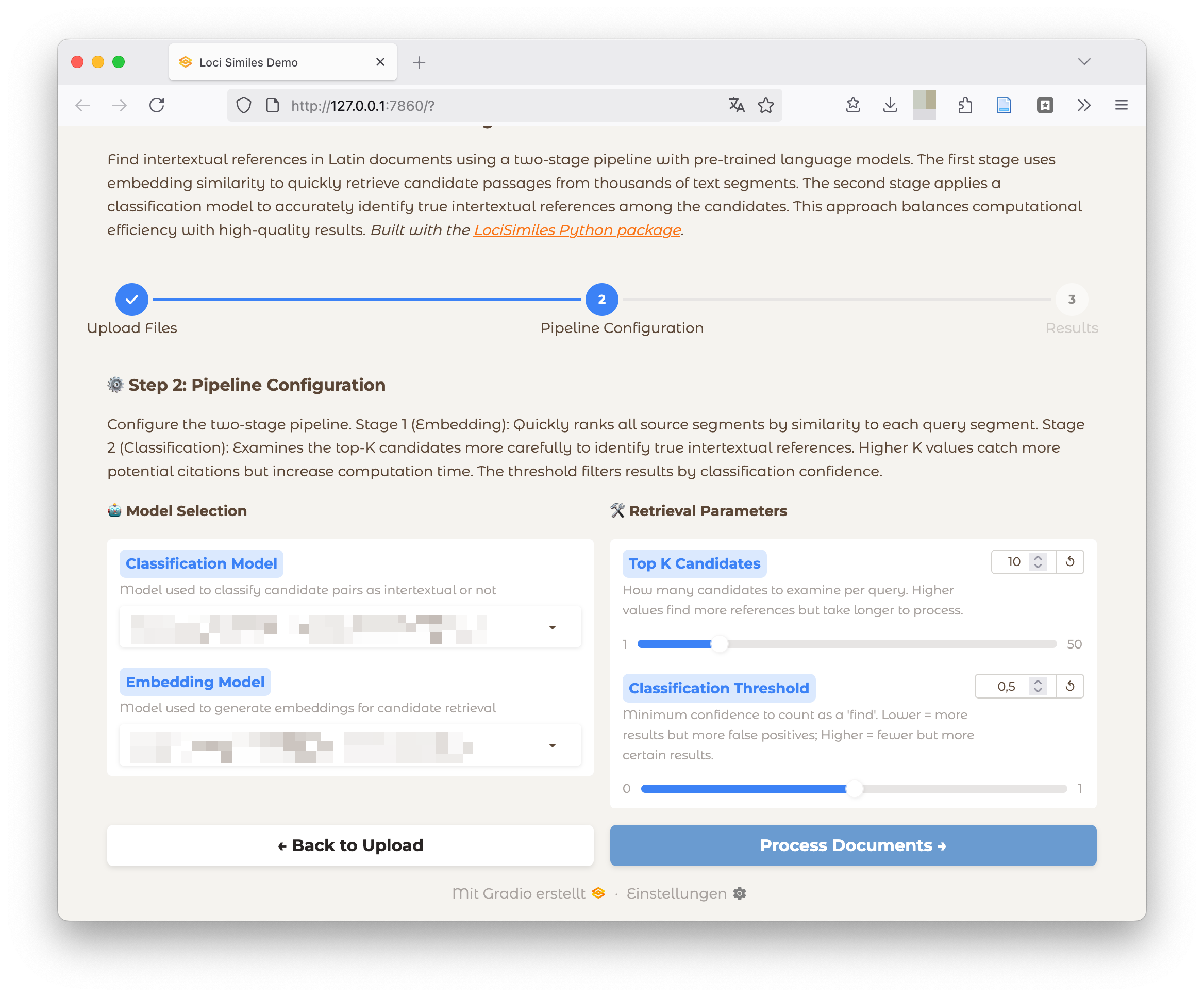}
    \end{subfigure}
    
    \vspace{0.5cm} 
    
    \begin{subfigure}[b]{1.0\textwidth}
        \centering
        \caption{Interactive Result Exploration}
        \label{fig:gui_results}
        \vspace{5pt}
        \includegraphics[width=\linewidth]{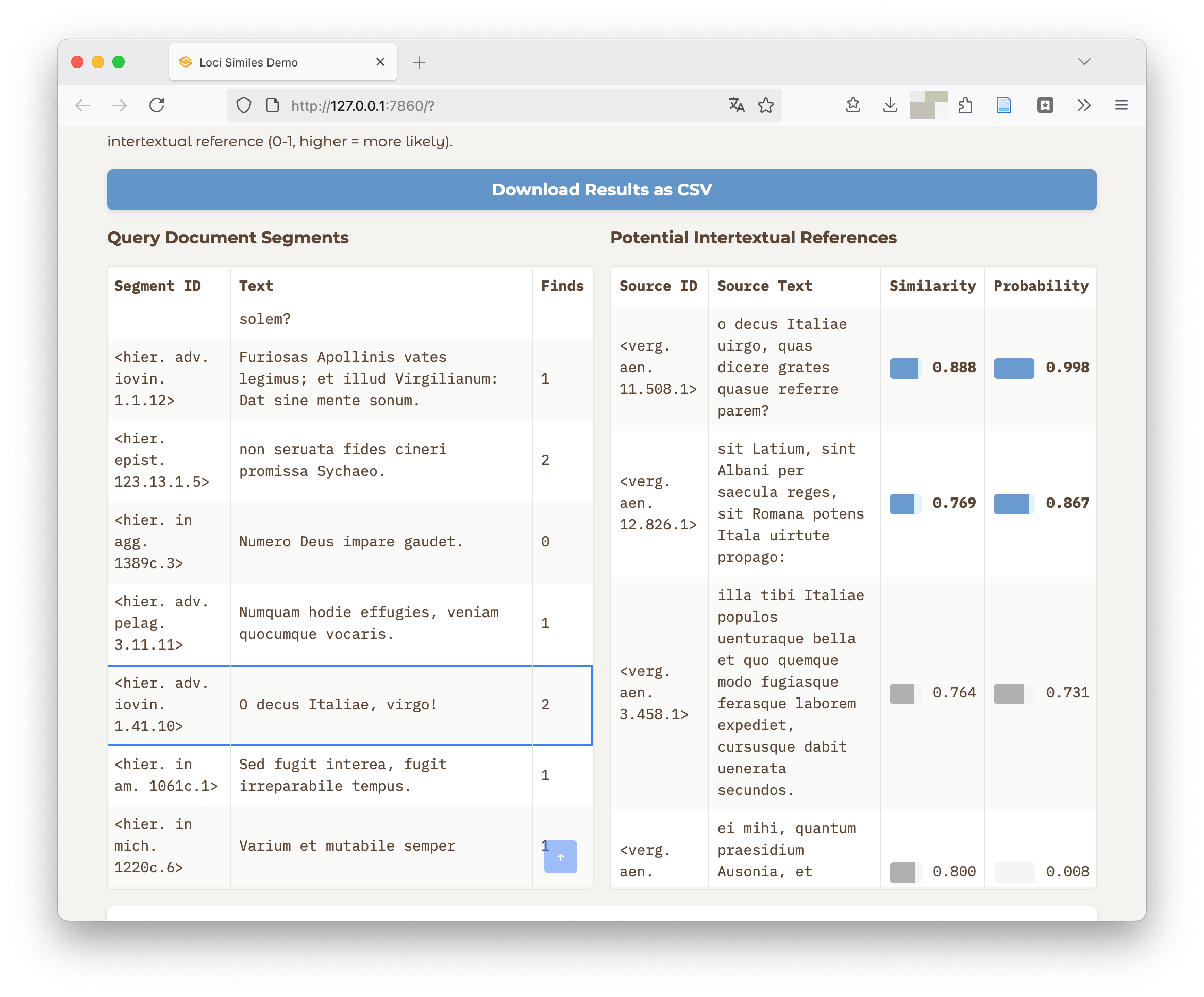}
    \end{subfigure}
    
    \caption{\textbf{Graphical user interface workflow.} \textit{(A) Data Upload:} Users ingest query and source documents via CSV files. \textit{(B) Configuration:} The pipeline is customized by selecting pre-trained models and tuning the retrieval depth (top-$k$) and classification confidence threshold. \textit{(C) Result Exploration:} The dashboard presents query segments alongside retrieved source candidates, displaying both cosine similarity and classification probability scores, with options to export matches as a CSV file.}
    \label{fig:gui_workflow_complete}
\end{figure*}

\paragraph{Workflow.} The application workflow is organized into three sequential stages, as illustrated in Figure~\ref{fig:gui_workflow_complete}:
\begin{enumerate}
    \item \textbf{Data Upload:} Users upload custom query and source documents as CSV files.
    \item \textbf{Configuration:} The pipeline is customized by selecting pre-trained models and tuning retrieval parameters (e.g., retrieval depth $k$ and classification confidence thresholds).
    \item \textbf{Result Exploration:} The interactive dashboard presents query segments alongside retrieved source candidates, displaying cosine similarity and classification probability scores, with functionality to export confirmed matches.
\end{enumerate}
The three panels show the upload view (Figure~\ref{fig:gui_upload}), configuration view (Figure~\ref{fig:gui_config}), and result-exploration view (Figure~\ref{fig:gui_results}).

%% file: sections/appendix/13_scalability.tex
\section{Scalability and Future Directions}
\label{app:scalability}

\paragraph{Scalability of the pipeline.} The full corpus considered here contains roughly $7.7 \times 10^{9}$ possible query--source pairs. The retrieval stage reduces this space by orders of magnitude, so end-to-end cost is dominated by (i)~the dense index built once per source corpus and (ii)~cross-encoder passes on the top-$k$ candidates per query. At larger corpus scales, neither component remains practical without dedicated indexing: dense retrieval must move from exact to approximate nearest-neighbor search, and the cross-encoder budget must be controlled by reducing $k$ or by precomputing partial token representations. \citet{DBLP:journals/ijdsa/MahadevanMMT25} discuss this billion-scale regime for historical text reuse and are the closest point of comparison for scaling \textit{Loci Similes}.

\paragraph{Completeness of the ground truth.} As noted in the Limitations, the expert-verified ground truth is not exhaustive. Some candidate pairs counted here as false positives may correspond to genuine but undocumented intertextual references, so precision figures on the curated subset should be read as lower bounds rather than as estimates over the complete reference graph.

\paragraph{Directions opened by the benchmark.} Beyond the modeling directions in Section~\ref{sec:conclusion}, the benchmark supports two further lines of work: generative explanation, where a model produces a natural-language rationale for a proposed reference that can be checked against expert annotations for faithfulness; and indexing-aware retrieval, where late-interaction retrievers such as ColBERT~\citep{DBLP:conf/sigir/KhattabZ20} and the data structures surveyed by \citet{DBLP:journals/ijdsa/MahadevanMMT25} are evaluated under the same staged protocol, reporting quality and cost jointly.